\documentclass[manuscript,screen,nonacm]{acmart}

\AtBeginDocument{%
  \providecommand\BibTeX{{%
    \normalfont B\kern-0.5em{\scshape i\kern-0.25em b}\kern-0.8em\TeX}}}

\AtBeginDocument{%
  \providecommand\BibTeX{{%
    \normalfont B\kern-0.5em{\scshape i\kern-0.25em b}\kern-0.8em\TeX}}}
    
\usepackage{algorithm}
\usepackage{xcolor}
\usepackage{algpseudocode}
\usepackage{multirow}
\usepackage{colortbl}
\AtBeginDocument{%
  \providecommand\BibTeX{{%
    Bib\TeX}}}

\usepackage{arydshln}
\definecolor{DarkGreen}{rgb}{0.0, 0.5, 0.0}

\newcommand{\pquote}[1]{\emph{``#1''}}

\newcommand{\aireply}[1]{``\texttt{#1}''}

\begin{document}

\title[Interaction Design with Generative AI]{Interaction Design with Generative AI: An Empirical Study of Emerging Strategies Across the Four Phases of Design}

\author{Marie Muehlhaus}
\email{muehlhaus@cs.uni-saarland.de}
\affiliation{%
  \institution{Saarland University, Saarland Informatics Campus}
  \city{Saarbrücken}
  \country{Germany}
}

\author{Jürgen Steimle}
\email{steimle@cs.uni-saarland.de}
\affiliation{%
  \institution{Saarland University, Saarland Informatics Campus}
  \city{Saarbrücken}
  \country{Germany}
}

\begin{abstract}
Generative Artificial Intelligence (Generative AI) holds significant promise in reshaping interactive systems design, yet its potential across the four key phases of human-centered design remains underexplored. 
This article addresses this gap by investigating how Generative AI contributes to requirements elicitation, conceptual design, physical design, and evaluation. Based on empirical findings from a comprehensive eight-week study, we provide detailed empirical accounts and comparisons of successful strategies for diverse design activities across all key phases, along with recurring prompting patterns and challenges faced. Our results demonstrate that Generative AI can successfully support the designer in all key phases, but the generated
outcomes require manual quality assessments. Further, our analysis revealed that the successful prompting patterns used to create or evaluate outcomes of design activities require different structures depending on the phase of the design and the specific design activity.
We derive implications for designers and future tools that support interaction design with Generative AI. 

\end{abstract}



\maketitle

\section{Introduction}
The user-centered design cycle is commonly modeled in four key phases, comprising requirement elicitation, conceptual and physical design, and evaluation (ISO 9241-210:2019). Thus far, human-centered design in large parts has been an activity performed by humans with humans, requiring considerable manual effort, time, resources, and methodological expertise. Recent advances in Generative Artificial Intelligence (Generative AI) hold promise to support the design process in ways that were solely attributed to humans so far. Of note, now widely available and easily accessible Large Language Models (LLMs) like ChatGPT~\cite{brown_arxiv2020}, Bard\footnote{Google Bard: \url{https://bard.google.com/chat} (accessed 02/05/2024)}, and Copilot\footnote{Microsoft Copilot: \url{https://copilot.microsoft.com/} (accessed 02/05/2024)}, encode both general and domain-specific knowledge that can be relevant for design processes, while powerful image-generation models like DALL-E~\cite{ramesh_arxiv2021} and Midjourney\footnote{Midjourney: \url{https://www.midjourney.com/} (accessed 02/05/2024)} hold promise for visual design tasks.

Recent work highlights the potential of Generative AI to reshape the traditional human-centered design process~\cite{schmidt_eics23,schmidt_interactions24}. It was suggested that Generative AI might complement the design process, extend the range of existing methods for human-centered design, and in parts might ultimately even replace costly human work~\cite{schmidt_interactions24}. 
Initial work started exploring Generative AI's potential for specific design activities, such as for persona creation~\cite{goel_chiwork23}, or generating synthetic user data~\cite{hamalainen_chi23}. 

However, only very little work has investigated how Generative AI tools can be used for interaction design from a holistic and more practical perspective. This aligns with prior work that highlights the difficulties of prompt design to steer the generated outputs and points to a lack of systematic research from an HCI perspective~\cite{dang_arxiv2022}. 
Specifically, we currently lack a comprehensive understanding of Generative AI's potential across the key phases of the design process, alongside a systematic investigation of successful usage strategies and comparison of how these strategies might be different or similar in different design phases. This is crucial to guide more targeted use of Generative AI in interaction design.

This article presents empirical insights from a systematic exploration of Generative AI's potential in the user-centered design cycle. These allow us to systematically compare different strategies and prompting patterns of Generative AI across the different phases of design.

We conducted an eight-week study (n=10) within a graduate-level human-computer interaction course, in which participants used GPT-4 and DALL-E3 to explore how Generative AI can support interaction design for a given design task.  
The participants engaged in diverse design activities in a systematic procedure, covering the four design phases, and documented successful and failed approaches of how they used Generative AI. We evaluate selected AI-generated outcomes or \textit{artifacts} through another user study (n=7), in which a new pool of participants was tasked to rate the quality, adequacy of provided detail, and relevance to the given design task.

A qualitative analysis revealed that Generative AI offered helpful support to designers across all phases of user-centered design, spanning such versatile activities as analyzing usage contexts, creating design solutions and performing critical assessments not only of design solutions but also of evaluation methods. 
For each of them, we report on successful strategies identified by our participants. These strategies cover versatile design activities across different design phases, for instance, demonstrating how Generative AI was used to create a set of elaborated personas or to evaluate an experimental design. We further extend beyond prior work by providing in-depth empirical accounts of successful prompting patterns, including their key elements, example inputs and outputs, and by discussing the quality of the provided results. We conclude by discussing remaining challenges and implications for designers, for future tools and for the role of Generative AI in interaction design. 

The key findings of this article are:
\begin{itemize}
    \item Generative AI can effectively support the designer across various design activities and phases to create artifacts for design activities from scratch, to iteratively refine them, or to gather feedback on existing artifacts. Yet, the generated artifacts require manual refinement. Based on these insights, we propose a novel meta-design cycle that interleaves rapid AI-driven iterations and human guidance (see Figure~\ref{fig:design_cycle}).
    \item We provide detailed empirical accounts and practical insights of successful examples for the requirements analysis, design and evaluation phases of user-centered design. We further discuss the AI-generated artifacts by evaluating their quality, the adequacy of the level of detail provided, and their relevance to the given design task.
    \item We identify recurring prompting patterns of successful strategies. Our results reveal that design activities in earlier design phases benefited from single and oftentimes highly structured prompts, whereas iterative prompting proved helpful for advanced design phases. To gather feedback on existing artifacts, participants deployed single prompts that provided minimal additional information. Persona patterns~\cite{white_arxiv2023} were deployed throughout various design phases, but with different purposes depending on the respective phase. 
    \item We identify challenges that our participants faced across the key phases of design, comprising stereotypical representation of users and scenarios in the early design phases, difficulties in generating functional designs and conveying technical or subjective information relevant to later design phases, and ethical concerns in using commercial Generative AI tools in interaction design.
\end{itemize}

This research offers practical insights into the use of Generative AI in user-centered design. We envision our work to enrich our understanding of the potential of Generative AI and promote more effective and innovative design practices in interactive systems--a relevant consideration, given the rapid advancement of Generative AI capabilities and the need for a timely understanding of its potential impacts and risks. 
Thereby, this paper is targeted at designers, developers of AI tools, and researchers interested in practical applications of Generative AI.
\section{Related Work}
This work is motivated by prior work on AI to support creativity and co-creation, the rise of Generative AI and LLMs, and their potential in interaction design.

\subsection{AI for Creativity Support and Co-creation}
    With the increasing capabilities of AI, new tools are emerging that not only enhance human creativity but also aid in creating artifacts across various domains. For instance, some tools assist in the ideation of stories through trope knowledge~\cite{chou_uist2023}, create storybooks~\cite{yan_uist2023}, facilitate fashion design~\cite{wu_chi2023,jeon_chi2021}, or co-create music~\cite{louie_chi2020}.
    Moreover, their use extends beyond traditional domains, such as aiding designers with 3D modeling~\cite{faruqi_uist2023,liu_dis2023}, producing video game scenes mimicking human directors' styles~\cite{evin_chiplay2022}, or simulating artificial agents that mimic human behavior~\cite{park_uist2023}. However, according to Frich et al., these tools primarily focus on generating and implementing ideas so far~\cite{frich_chi2019}. Less common are tools that assist in evaluation and identifying problems. Hence, recent work prompts researchers to investigate the role of AI for more complex tasks that not only aid in automating but augment human creativity~\cite{ding_arxiv2023}. 
    Our work adds to this body by investigating how Generative AI can support designers to create new and evaluate existing artifacts of various design activities and phases of interaction design.

\subsection{Generative AI, Large Language Models, and Prompt Engineering}
Generative AI, and in particular LLMs, have become increasingly popular due to widely available tools like ChatGPT, Bing, and Bard. LLMs are deep learning models capable of understanding and generating natural language. 
What sets LLMs apart is their ability to learn from huge amounts of data from various sources such as forums, websites, scientific literature, and books. With their extensive knowledge across different domains, LLMs have become valuable tools for diverse usage. For example, they can simulate characters in role-playing scenarios~\cite{shanahan_nature2023}, allow users to style websites through natural language~\cite{kim_chi2022}, or help writers create scientific metaphors~\cite{kim_dis2023}. 

While LLMs strive to facilitate natural interactions akin to human language, understanding how to effectively prompt them is essential. A growing body of literature tackles the challenge of prompt engineering from different perspectives. Some works formalize and categorize prompts~\cite{liu_arxiv2021}, while others share best practices or guidelines~\cite{zamfirescu_chi2023, white_arxiv2023, ekin_authorea2023}. For instance, researchers have compiled catalogs of prompting patterns, offering diverse strategies for interacting with LLMs, such as prompting the AI to assume a character's role or prompting the AI to ask the user questions~\cite{white_arxiv2023}. However, designing effective prompts by combining these elementary patterns demands considerable effort and only few works systematically explored prompting within the field of HCI~\cite{dang_arxiv2022}. One notable example in this area explored the effects of prompt keywords and model hyperparameters on generated images at a large scale, deriving design guidelines for prompt engineering in the context of text-to-image Generative AI~\cite{liu_chi2022}. A more recent seminal work employed design probes to investigate how non-experts approach prompt design and identify barriers~\cite{zamfirescu_chi2023}. 
Our work contributes to this body of research through detailed empirical accounts and practical insights into the usage of Generative AI in the area of interaction design. 

\subsection{(Generative) AI in Interaction Design}

Two different research directions deal with AI in interaction design: One focuses on how to design \textit{for} AI. For instance, prior work investigated how UX practitioners approach the design of AI-based products, identifying emerging design patterns~\cite{windl_chi2022} and adaptations of UX practices~\cite{wang_chi2023}.
The other direction focuses on how to design \textit{with} AI to enhance the design process and the outcome.   
Computational approaches have a long tradition in this field. For instance, prior work demonstrated that these approaches help designers with ideation~\cite{koch_chi2019} or optimizing UI and sensor layouts of interactive systems~\cite{bailly_uist2013,oulasvirta_ieee2020,nittala_nature2021}. Recent work has opened up a discussion about the significance of Generative AI, and particularly LLMs, as a new means to enhance the design process~\cite{schmidt_eics23,schmidt_interactions24} and delved into different aspects of interaction design, exploring how specialized tools aid in tasks like requirements analysis or evaluation~\cite{zhang_ieee2023,kuang_chi2023,gebreegziabher_chi2023}. But the potential of general-purpose Generative AI and LLMs for the different phases of interaction design has been hardly investigated so far. 
A limited number of works have examined the process of eliciting requirements, with particular focus on generating personas~\cite{goel_chiwork23,zhang_ieee2023}. For instance, Goel et al. demonstrated that design novices can create satisfactory personas with the help of GPT-3 and argued that detailed prompts were more likely to lead to a successful persona~\cite{goel_chiwork23}. A more comprehensive investigation identified opportunities and challenges of LLMs for eliciting, analyzing, specifying, and validating software requirements~\cite{arora_arXiv23}.
In the design phase, prior work highlights Generative AI's principled potential to aid rapid design ideations, but also the system’s lack of understanding the context of the design task~\cite{tholander_dis23}. Another work in this area explored co-ideation with image generators, investigating how designers and image generators could collaborate in the future~\cite{chiou_dis2023}. This work provides practical insights into prompting patterns, with one of the key findings centering around how the level of detail influences the preciseness of generated outcomes. Recent work further explored the potential of LLMs to augmenting design tools for physical devices~\cite{lu_arXiv24}.
In the evaluation phase, recent research demonstrated GPT-3's potential to generate synthetic research data for HCI~\cite{hamalainen_chi23}.
These works provide valuable insights into the potential of Generative AI for specific design activities. Our work extends this body of research through in-depth empirical accounts and a systematic comparison of successful usage strategies across the key phases of design. This is a crucial consideration to guide more targeted use of Generative AI in interaction design.

\section{Method}
To better understand the potential of Generative AI in interaction design and to gain practical insights, we conducted two user studies in which participants (1) systematically explored how they can integrate Generative AI into various design activities and (2) assessed the quality of the resulting AI-generated artifacts.

\subsection{Design Task}\label{subsec:task}
Inspired by~\cite{tholander_dis23}, we formulated a design task for the first user study, which comprised a design challenge, situation of concern, and problem statement. It was created based on the outcomes of a published research project~\cite{saberpour_uist2023} which presented a wearable robotic limb able to create haptic feedback on diverse body parts. The participants were provided with the following design challenge: \textit{``You have been asked to improve the gaming experience of VR applications, with a particular focus on the haptic experience''}. Further, the participants were informed about the intended solution, which should consist of a novel device that is wearable and can create a sense of touch at different locations of the user’s body without restricting the user’s mobility. The detailed description of the task can be found in Appendix~\ref{sec:task_description}. The task was selected due to its versatility, as it also covers aspects like the physicality of design, e.g., haptic experiences and hardware design. 

We assigned each participant of the first user study a specific design activity (see Table~\ref{tab:participants}) and asked them to explore different strategies of how Generative AI could be leveraged to contribute to the success of this design activity. The provided design task hereby serves as an example application that guides participants in creating specific textual prompts that triggered the AI to generate the desired textual or visual artifacts.
To address the fact that participants worked on the four design phases in parallel, we additionally provided them with fictive outcomes from the preceding design phases: Those participants who worked on a design activity belonging to the conceptual or a later design phase received a list of stakeholders, two personas and a scenario, the physical design and evaluation groups further received conceptual sketches and a storyboard, and the evaluation group a description, photos and a video of the final prototype taken from~\cite{saberpour_uist2023}.

\paragraph{Exploration Structure}
Participants were tasked to explore each strategy following a systematic procedure. 
The five steps of the procedure were synthesized from similar procedures presented in prior work~\cite{zamfirescu_chi2023, white_arxiv2023} and resemble an iterative design cycle: 
\begin{itemize}
    \item \textbf{Identify the strategy's core idea}: What is the designer's problem that this strategy solves and which goals does the strategy achieve? Why is it important to solve this problem?
    \item \textbf{Identify the strategy's high-level functionality}: What is the functionality of the Generative AI? What are the input and output modalities? 
    \item \textbf{Design the prompt and interaction structure}: What key information does the input (prompt) of the strategy provide to the Generative AI? And how is this information provided?
    \item \textbf{Evaluate the strategy}: Assess the strategy's practical capabilities by testing several examples. The generated examples should be related to the design task. Based on the examples, what are frequent errors and how critical are they? What are the successes of the strategy?
    \item \textbf{Iterate on the strategy}: Similar to an optimization process, pick one of the identified errors and iterate on the strategy to resolve it.
\end{itemize}

\begin{table*}
\begin{tabular}{ p{1.75cm}||p{3cm}|p{8.25cm}  }
 \hline
 \textbf{Design Phases} & \textbf{Design Activity \& Participant ID} & \textbf{Participant's Experience} \\
 \hline
 \multirow{2}{1.75cm}{Requirement analysis}  & Create personas (P1) & background in HCI and LLMs through a student job; personal experience using and implementing LLMs.\\
 \cdashline{2-3}
  & Create scenarios (P2) & background in HCI and Generative AI through university courses\\
  \hline
 \multirow{3}{1.75cm}{Conceptual design} & Create conceptual ideas (P3) & background in NLP and deep learning through university courses; personal experience using LLMs\\
 \cdashline{2-3}
 & Assess conceptual ideas (P4) & background in HCI through university courses; personal experience using LLMs\\
 \cdashline{2-3}
  & Design haptic experiences (P5) & professional background in HCI, haptic design, and VR; limited personal experience using LLMs\\
  \hline
 \multirow{3}{1.75cm}{Physical design} & Design a prototype (P6) & background in HCI, NLP, and game technologies through university courses; limited personal experience using LLMs\\
 \cdashline{2-3}
  & Select components (P7) &  professional background in HCI and fabrication\\
 \cdashline{2-3}
  & Program the hardware (P8) & background in HCI and Generative AI through university courses; personal experience using LLMs \\
 \hline
 \multirow{2}{1.75cm}{Evaluation} 
  & Identify the study method (P9) & background in HCI through university courses\\ 
 \cdashline{2-3}
  & Design a controlled experiment (P10) & background in HCI through university courses and student jobs; personal experience using LLMs\\
  \hline
\end{tabular}
\caption{The ten selected design activities in the four key design phases. Every design phase was worked on by a group of two to three participants. Each participant engaged in a specific design activity.}
\label{tab:participants}
\end{table*}

\subsection{Procedure}
\paragraph{Exploring the usage of Generative AI in interaction design}
The first study was conducted as part of a graduate-level research seminar on human-computer interaction, for a duration of 8 weeks. The tasks involved 8-10 hours of work per week for each participant. 
The study was divided into four phases:

In the first week, we provided participants with background literature on prompt engineering~\cite{zamfirescu_chi2023, white_arxiv2023, ekin_authorea2023} to mitigate initial selection and use barriers~\cite{zamfirescu_chi2023,ko_ieee2004}. Further, we assigned participants to groups, each working within one of the four design phases (requirements, conceptual design, physical design, or evaluation).

The second week comprised an initial exploration of interaction design with Generative AI. We introduced the structure of the systematic exploration (see above) as part of a 2-hour plenary session and participants gathered first hands-on experience through a practical exercise, with the opportunity to clarify open questions. Furthermore, each group was provided with the design task (detailed on in Subsection~\ref{subsec:task}) and then explored a provided list of design activities in order to identify particularly promising ones to proceed with. 

In the third week, participants selected a specific design activity (see Table~\ref{tab:participants}) and started to explore addressing it with Generative AI. Weekly group meetings with an assigned advisor facilitated exchanging ideas and giving feedback between the group members. The groups submitted written documentations of their exploration at the end of week six. 

In the last two weeks, group members discussed their individual findings, synthesized them into principled opportunities, challenges, and recommendations for using Generative AI in their assigned design phase, and submitted another written report.
Finally, they were informed that they have the opportunity to provide their data for scientific analysis, on a voluntary opt-in basis. Those who decided to opt in signed a consent form and provided their demographic information.

\paragraph{Assessing the quality of explored strategies}
We then assessed the quality of the strategies explored in the first study. Hereby, we provided new participants with eight selected examples that participants of the first study had generated. These examples consist of sequences of prompts and AI-generated outputs (see Appendices~\ref{app:persona_generation}--\ref{app:modify_experiment}) and are introduced in more detail in Section~\ref{sec:overview}. For each provided example, participants were tasked to rate the quality of the generated output, based on the input information that the designer provided. 
We assessed the quality of the artifacts through a questionnaire that consists of a 7-point Likert scale with three items: \pquote{The quality of the generated artifact is as good as what an experienced designer could have achieved}, \pquote{The generated artifact has an adequate level of detail (neither missing nor irrelevant details)}, \pquote{The generated artifact is relevant to the design task}.
For each item, participants were requested to provide additional qualitative feedback on their ratings.

\subsection{Tools}
In the initial exploration phase of the first user study, participants used OpenAI's ChatGPT 4~\cite{brown_arxiv2020}, but we changed to Bing Chat (now \textit{Copilot}) for the main exploration phase due to access issues (see Section~\ref{subsec:challenges} for details). At the time of the study, as of December 2023, Bing Chat is powered by a transformer-based language model which is based on GPT-4 and uses Bing Image Creator powered by DALL-E3 to generate images.  

\subsection{Participants}
Ten participants (3 f, 6 m, 1 d) took part in the first study (P1 - P10). Two were in the PhD preparatory phase in HCI, seven were Master students of Media Informatics, Computer Science and Visual Computing, and one was a Bachelor student in Media Informatics. Table~\ref{tab:participants} provides an overview of their experience with HCI and Generative AI.

For the second user study, we recruited seven new participants (3 f, 4 m, 0 d) from our academic network (P11 - P17). Our eligibility criteria were participants with knowledge in human-centered design, gained through research and/or teaching activities in HCI-related courses at university. Participants’ backgrounds included Master and PhD students as well as a PostDoc in Human-computer Interaction.

\subsection{Data Analysis}
We employed an inductive and iterative approach to analyze the submitted documents.
First, the first author familiarized herself with the data, then conducted two qualitative analyses for distinct objectives: One focused on identifying patterns in how the participants prompted Generative AI for different design activities across design phases. The other analysis focused on identifying the successes and limitations of AI-generated artifacts, alongside principled challenges the participants faced. 
The first author identified these patterns and themes through an iterative process of clustering, splitting, and merging. This process was intermittently complemented by discussions of the identified patterns and themes among the co-authors and repeated until a consensus was reached.

\subsection{Positionality}
All co-authors and advisors of the participant groups have a background in Computer Science and Human-Computer Interaction. Our advising approach aimed to maintain objectivity and avoid steering the groups in certain directions. The first author was involved in advising the groups and conducted the data analysis. 
Further, it is important to note that since the study took place as part of a university seminar, there is an instructor-student dependency and the students' performance impacts their final grade. We clearly communicated that opting in or out of the user study and the follow-up data analysis will neither affect the admission to the course, the course procedure, nor the final grading in any way.

\section{How Generative AI was Used for Interaction Design Activities}\label{sec:overview}
\begin{figure*}[p]

\centering
\includegraphics[width=1.4\textwidth, angle =90 ]{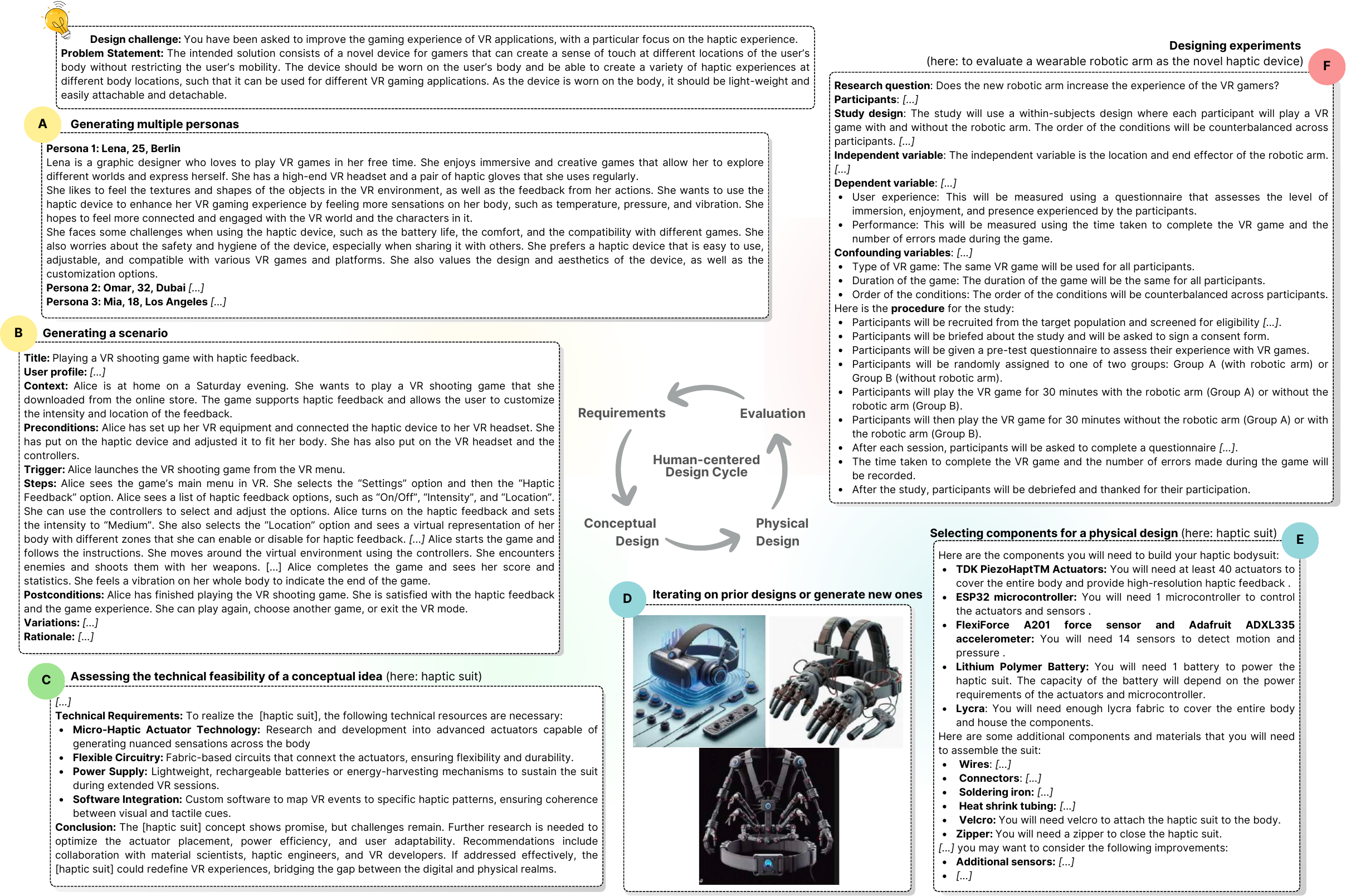}
\caption{An overview of artifacts generated by the study participants. The artifacts cover selected examples and demonstrate how Generative AI could make helpful contributions in the four phases of user-centered design.}
\label{fig:awesome_image}

\end{figure*}

The results of our study show that Copilot could make helpful contributions in all phases of interaction design. The AI helped in identifying requirements in the early design phases by generating personas and scenarios; it contributed to the creative design phase by refining conceptual designs and assessing their potential; it assisted in the implementation of prototypes by identifying suitable hardware components; and it finally helped in designing evaluation studies, by proposing designs for controlled experiments and by identifying problems with experimental designs. This versatility in the support provided -- comprising such diverse activities as analysis of the usage context, creation of design solutions, and critical assessment both of designs and methods -- makes AI a promising tool to assist interaction designers throughout the entire design process. 

In the following, we start by presenting an overview of how participants have used Copilot for interaction design, before we analyze further below in more detail the prompting patterns and the quality of the generated results. We prioritized those examples that provided successful\footnote{We consider examples 'successful' if they provided reasonable outputs which avoid overly generic, wrong, or unhelpful results, and have been indicated as successful by the participants.} outcomes or \textit{artifacts} and were thoroughly documented by the participant. We now provide empirical accounts, structured by the key phases of user-centered design:

\subsection{Requirements Analysis}
In the requirements analysis phase, participants have used Copilot to get support with generating personas and scenarios -- both essential methods for identifying the requirements of a product~\cite{cooper_wiley2014}.
\textit{Personas} represent the product's stakeholders. Since a product usually has a diverse set of stakeholders, creating multiple personas is useful for designers to capture the stakeholders' differing needs and preferences. As the established practice is to generate personas manually, this can be time-consuming. P1 explored how Copilot can speed up the activity by rapidly generating a set of personas for the provided design task. 
Figure~\ref{fig:awesome_image}A depicts a representative excerpt of output; more can be found in Appendix~\ref{app:persona}. Examples of generated personas encompass: the graphic designer \aireply{Lena, 25, Berlin} who \aireply{enjoys immersive and creative [VR] games} in her free time and wants \aireply{a haptic device that is easy to use, adjustable, and compatible with various VR games and platforms. She also values the design and aesthetics of the device}; and the software engineer \aireply{Omar, 32, Dubai} who \aireply{likes to play VR games as a way to [\ldots] socialize with his friends online. He prefers competitive and cooperative games that challenge his skills and strategy} and \aireply{prefers a haptic device that is fast, precise, and [\ldots] that can provide a wide range of feedback on different body locations}. 
While we will discuss the quality of the generated artifacts in more detail below, these examples show that the generated personas exhibit different character traits, challenges, and preferences relevant to the design task under consideration. Therefore, they possess important qualities that so far have been attributed to personas that were manually designed by experts.  

To better understand how a persona would interact with the device, P2 aimed to generate \textit{scenarios} for the given design challenge. Figure~\ref{fig:awesome_image}B depicts a representative excerpt of output; more can be found in Appendix~\ref{app:scenario}. In the provided example, Copilot suggests a scenario for \aireply{Alice [who] is a 25-year-old gamer who enjoys playing VR games for fun and relaxation. [\ldots] Alice is at home on a Saturday evening. She wants to play a VR shooting game that she downloaded from the online store.} Once this context has been provided, the generated scenario details the essential steps relevant to the interaction with the device. These include Alice, who, for instance, first connects the device, then customizes the intensity and location of the haptic feedback to her desires, before starting the game with the reconfigured haptic feedback. This information is useful as it points the designer to relevant functional requirements that specify what the novel device should do.   

\subsection{Conceptual and Physical Design}
In the conceptual design phase, it is common practice to generate many design alternatives and then identify the most promising designs for further refinement~\cite{preece_wiley2015}. The participants have used Copilot to assist in several of these activities: by assessing the potential of conceptual designs to aid comparison and by further refining conceptual designs. P3-P8 investigated a range of potential conceptual designs for the to-be-designed haptic device and the types of haptic feedback it might provide. Informed by fictive personas and scenarios from the requirements phase, these conceptual designs were either first manually identified by the participants or generated with the help of Copilot. Their design propositions included haptic gloves, suits and vests, adhesive skin patches, or wearable robotic limbs, each of which could enable different types of haptic feedback, such as vibrotactile and force feedback.

P4 then used Copilot to automatically generate assessment reports on selected designs in order to assess their potential, so he could identify a most promising one more rapidly. 
The participant's strategy was to indicate specific attributes, such as user experience, technical feasibility or aesthetics, based on which the concepts one after the other shall be assessed. Copilot should then generate detailed assessment reports for each concept, which the designer can then use for a manual comparison. 

Figure~\ref{fig:awesome_image}C provides an excerpt of an example output that illustrates a report that assesses the technical feasibility of a full-body haptic suit; more details can be found in Appendix~\ref{app:evaluate_concept}. In the example, Copilot points to relevant technical needs that the designer must consider, such as flexible circuitry and a fabric that prevents overheating of components. 
It further points to relevant technical challenges in the realization as \aireply{further research is needed to optimize the actuator placement, power efficiency, and user adaptability}. These are crucial aspects since poorly placed actuators and a lack of possibilities to adapt the feedback to user's personal needs and preferences can lead to an unsatisfactory user experience -- relevant challenges that the designer must consider.

Refining conceptual designs is an iterative activity, where new designs are created by modifying and enhancing existing ones. 
P6's goal was to design a wearable robotic device that generates haptic feedback. She decided to rapidly create a new design based on existing designs of related devices, here from devices she found on the Internet, by fusing them into a design that fulfills the requirements. 
As input, she provided images of existing haptic feedback devices, and tasked Copilot to identify common features of these devices and fuse them either into one or multiple images of a design that unifies the depicted concepts and that meets the specified requirements. We provide an example output in Figure~\ref{fig:awesome_image}D and the complete sequence of prompts and generated outputs in Appendix~\ref{app:modify_hardware}.
In the example, P6 provided as input four photos of prototypes -- all wearable devices that are strapped to users' hands and comprise robotic structures that create haptic feedback. 
Copilot fused these inputs and generated several design alternatives. One of these consists of \aireply{a haptic device for VR games that consists of a belt and a set of robotic arms}. The suggested design demonstrates a novel and creative concept for a haptic feedback device, that coarsely fuses the provided input images and aligns with the specified design requirements. It therefore serves as a useful source of inspiration for the designer.

Physical design differs from earlier design phases in that it requires making technically correct choices. P7 used Copilot in the physical design of a hardware prototype. Her goal was to identify which electrical components are needed. 
She adopted a co-creative approach to cater her needs for flexible and detailed decision-making. 
Figure~\ref{fig:awesome_image}E depicts an excerpt of the components for a full-body haptic suit that she selected together with the AI; the complete sequence of prompts and AI-generated replies can be found in Appendix~\ref{app:component_selection}.
The output comprises required components and suggestions, such as specific models for the microcontroller, sensors, actuators, and basic utilities like LiPo batteries, wires, textile, and a zipper, as well as rough estimations of the components' quantities. For instance, Copilot suggested the designer \aireply{will need at least 40 actuators to cover the entire body and provide high-level feedback}, and \aireply{14 [force sensors and accelerometers] to detect motion and pressure}. Outlining the most essential components alongside specific suggestions for the components and quantities to build a haptic suit, this strategy provides the designer with a valuable starting point for prototyping.

\subsection{Evaluation}

For the evaluation phase of user-centered design, participants have used Copilot to design controlled experiments as well as to identify problems with an experimental design of an evaluation study. 
In our provided design task, the final outcome of the prior design phases is a wearable device that provides haptic feedback using a robotic limb (see~\cite{saberpour_uist2023}).
Investigating the use of Copilot for the evaluation phase, P10 intended to design a controlled experiment in order to evaluate whether the device's haptic feedback can improve the VR experience. 
Figure~\ref{fig:awesome_image}F depicts a representative excerpt of output; more details can be found in Appendix~\ref{app:experiment_design}. 
The results demonstrate that after a single round of questions asked by the AI, the AI suggested independent and confounding variables as well as pointed to measures for the dependent variables.
For instance, for the dependent variables, it recommended to use \aireply{a questionnaire that assesses the level of immersion, enjoyment, and presence experienced by the participants} and measure the \aireply{Performance: This will be measured using the time taken to complete the VR game and the number of errors made during the game}. 
In addition, the AI suggested a step-by-step description of the study procedure that comprises a within-subjects design with counterbalanced conditions. While not highly detailed, the generated information provide a relevant outline of the overall study which the designer can iteratively refine. 

Next, P10 used Copilot to identify problems in the controlled experiment design. 
In the example provided in Appendix~\ref{app:evaluate_experiment}, P10 aimed to design a within-subjects study where participants play a VR game with and without the wearable device, to evaluate whether it improves the VR experience. The participant deliberately omitted information about counterbalancing. 
Copilot correctly identified that the study risks suffering from \aireply{order effects}, as the conditions were not counterbalanced. In addition, it pointed to other relevant aspects for re-consideration, such as suggesting to use an established questionnaire to measure immersion in VR. With this, it can help particularly inexperienced designers to rapidly improve experimental designs and uncover major flaws.

Finally, P10 aimed to use Copilot to further enhance the study design based on insights gained in a pilot study.
In the example provided in Appendix~\ref{app:modify_experiment}, P10 summarized the results of a fictive pilot study, stating that participants showed undesired learning effects across trials that impacted the measurements. Copilot successfully suggested improvements to address the described problem, such as to \aireply{providing participants with a tutorial or practice session before the study begins} to mitigate the initial learning effects and \aireply{ensuring that your participants are representative of [the] target population}. It further provided more recommendations such as to \aireply{consider using established measures [\ldots] and controlling confounding variables} and to \aireply{increase the number of trials}. These recommendations can be universally applied to diverse study designs beyond this specific example. With these suggestions for improvement, the AI cannot only point to problems in the study design, but also provide recommendations which can guide designers in iteratively refining their study designs.

\section{Prompting Patterns}
We next analyze the prompting patterns used to generate the artifacts presented in the previous section. Our results reveal that successful prompts require different structures depending on the phase of the design and the specific design activity: Single oftentimes highly structured prompts prove particularly helpful to generate artifacts for early phases of user-centered design, whereas creating solutions for advanced design phases benefited from iterative prompting. When gathering feedback on the quality of artifacts produced during design activities, single prompts with a less explicit structure were deployed. Persona patterns were used throughout the design phases, but with different purposes. 

\subsection{Single Prompts} 
Using single prompts appeared particularly suited for early phases of user-centered design to generate artifacts for design activities. 
This is because in earlier design phases, the designers had to consider less amount of details for input compared to later phases and hence could provide the essential information in a single prompt.

In particular, prompts that clearly define the structure of the expected output have proven useful as they help to ensure a consistent level of detail and quality in the generated outputs. 
For example, in the case of persona and scenario generation, P1 and P2 first tried to generate personas and scenarios using single and undetailed prompts, such as \textit{``Could you help me generate 10 personas for the following design challenge?''}, and then provided the design task. The generated artifacts, however, lacked diversity and a comparable level of detail, while also not sufficiently fulfilling the common structure and quality that is expected from personas or scenarios. 
Appendix~\ref{app:persona_generation_creative} shows an example. 
Using structured prompts, both P1 and P2 could overcome these problems and considerably enhance the quality of the artifacts (cf., Appendices~\ref{app:persona_generation} and~\ref{app:scenario}). 
The specific structure of the successful prompts, however, varies depending on the design activity. 
They are structured and comprise between 3 and 6 elements. Both had in common to specify the purpose (scenario or persona generation), context of the project (design challenge and intended solution) and the desired semantic information that the output should cover. For the persona, this involves, e.g., age, name, VR experience, and motivation of the stakeholder; for the scenario e.g., a description of the user, the user's physical context, and the interaction with the device. In addition, P1 also requested a creative and engaging tone and provided information about the desired syntax of the output (e.g., to start each persona with a number and name), which made for more detailed and consistent results. 

\begin{figure*}
        \centering
        \includegraphics[width=0.9\textwidth]{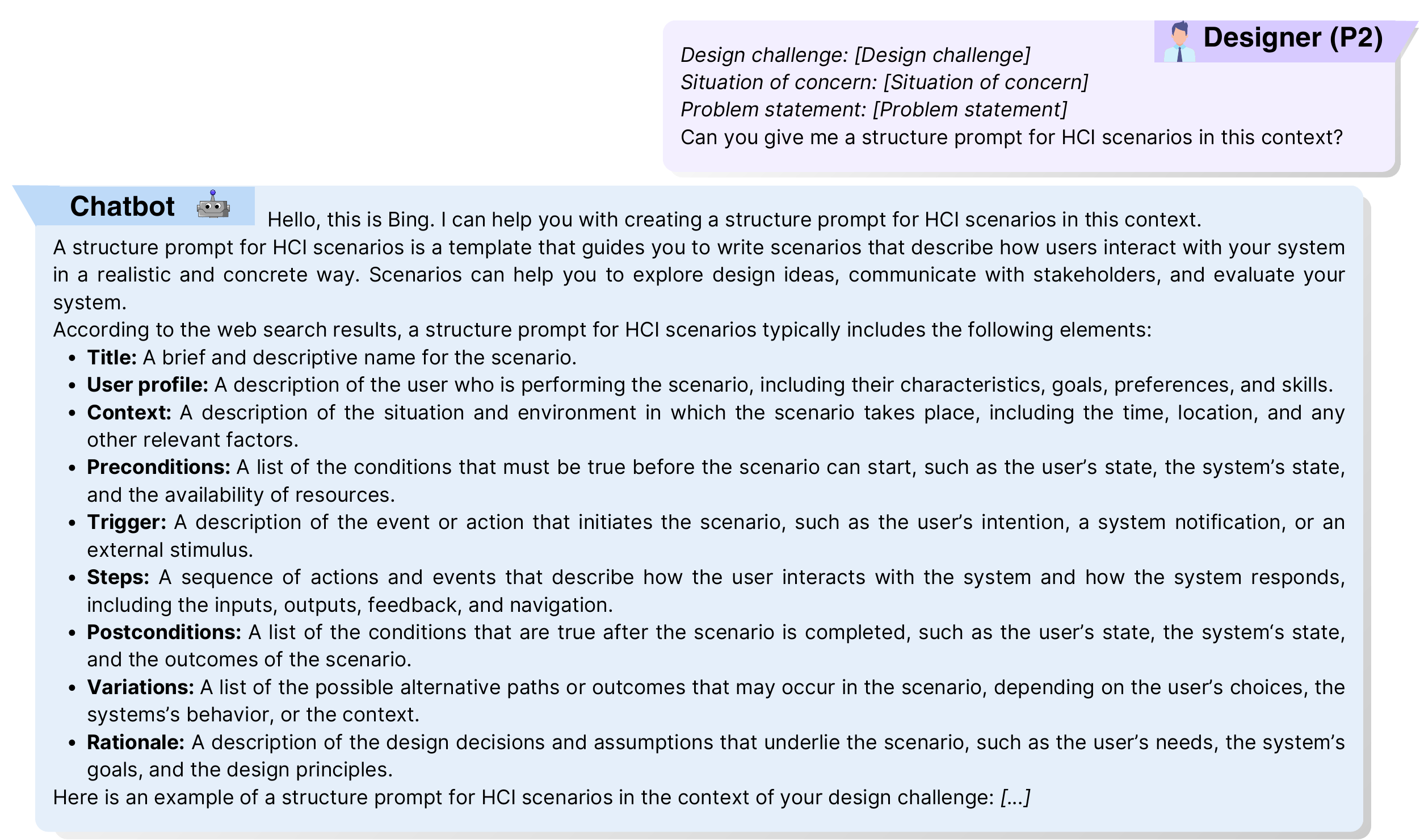}
        \caption{P2 used a meta-prompt to generate a structured prompt that he could use to create scenarios.}
        \label{fig:scenario_prompt}
\end{figure*}
To come up with these structured prompts for generating personas and scenarios, meta-prompting proved helpful. Meta-prompting is an approach that uses prompts to generate new prompts. Both P1 and P2 have used meta-prompts to inform their actual prompt. The meta-prompts covered (1) the purpose (generating a system prompt for persona/scenario generation), e.g., \pquote{I want to have general system prompt that I can use to make generate personas for the attached design challenge. Please include all the relevant details from the scenario to make the prompt as precise as possible}. Secondly, they covered (2) the context of the project, for which both P1 and P2 only provided the given design challenge, situation of concern, and problem statement. An example of a prompt used to generate scenarios is provided in Figure~\ref{fig:scenario_generation}.
P2 further pointed out that the use of the meta-prompting strategy does not only serve to improving the quality and diversity of the generated outputs. The result of the meta-prompt itself can also serve as a guide for an inexperienced designer, providing a scaffold that makes the structure of a design activity explicit and thereby helps the manual design process. For instance, the structure provided for the scenario generation (cf., Figure~\ref{fig:scenario_generation}) makes required aspects (such as \aireply{Trigger: A description of the event or action that initiates the scenario, such as the user’s intention, a system notification, or an external
stimulus.} or \aireply{Context: A description of the situation and environment in which the scenario takes place, including the time, location, and any other relevant factors}) explicit in a way that -- based on our experience -- is even hardly taught as such in HCI-related courses.

\subsection{Iterative Prompting}
\begin{figure*}
        \centering
        \includegraphics[width=\textwidth]{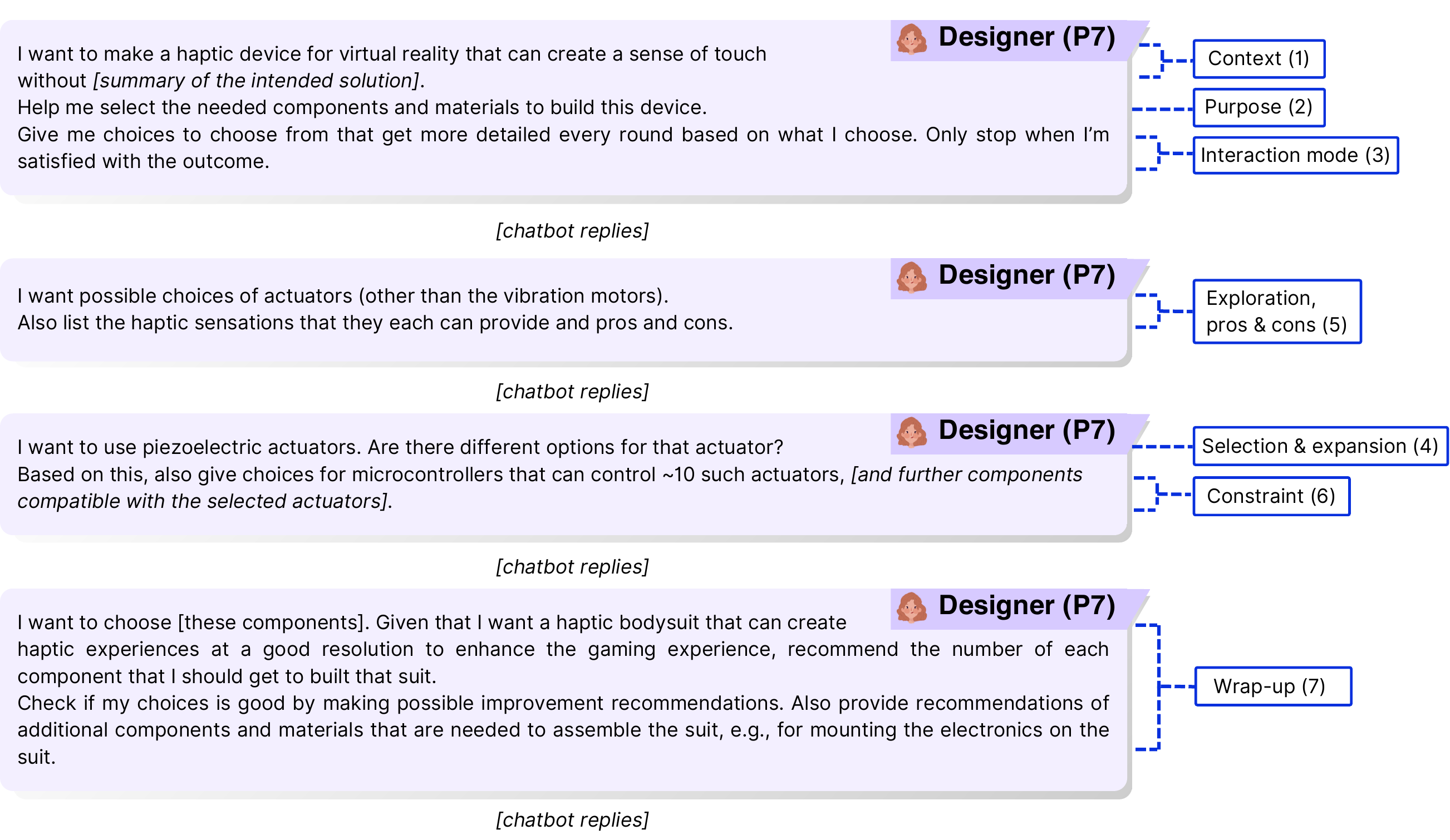}
        \caption{Prompts provided by P7 to identify components for a hardware prototype together with Generative AI. The key elements of this iterative prompting approach are annotated as explained in the text.}
        \label{fig:component_selection}
\end{figure*}

\begin{figure*}
        \centering
        \includegraphics[width=\textwidth]{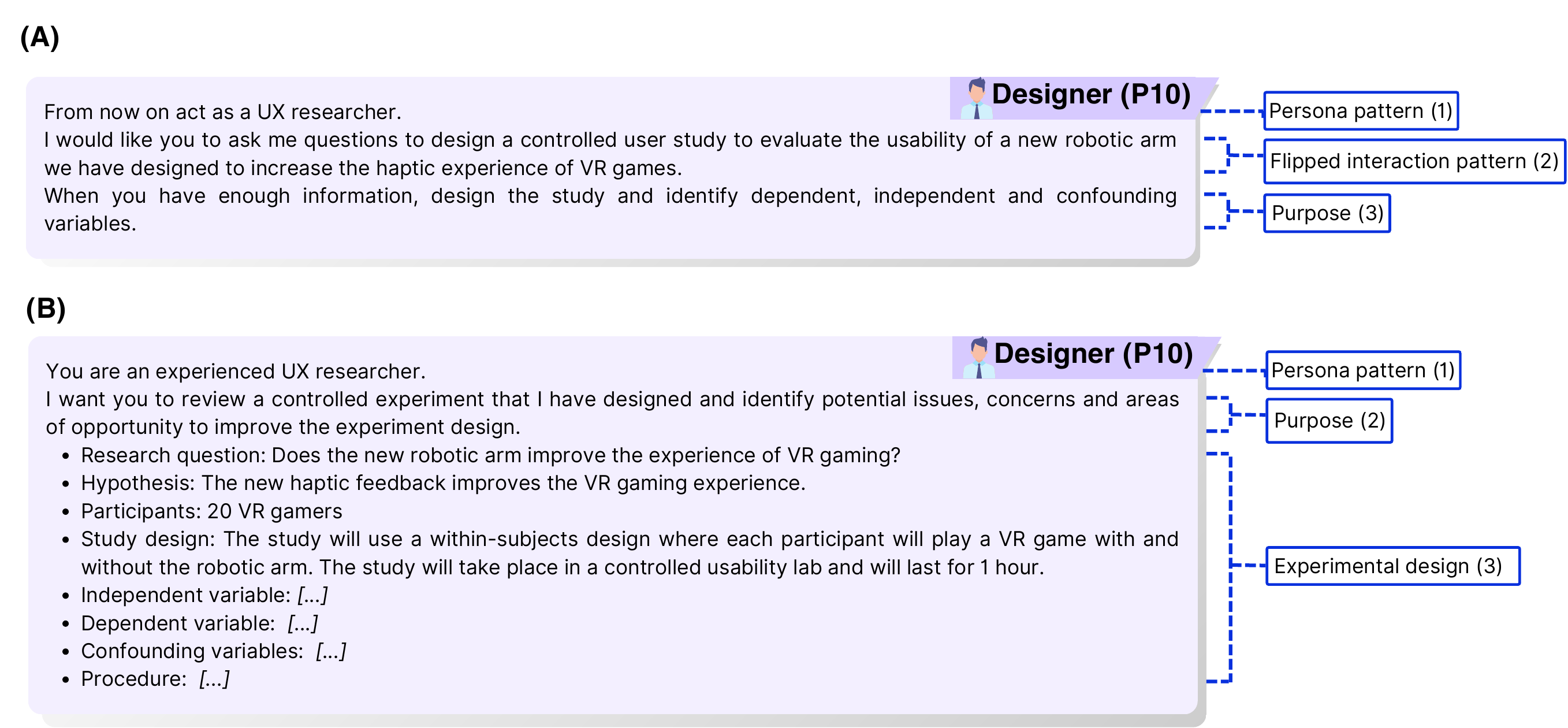}
        \caption{Two different prompting patterns deployed by P10. (A) P10 deployed a persona and a flipped interaction pattern to design a controlled experiment with Generative AI. The flipped interaction pattern initiates a conversation with the Generative AI. (B) A single prompt in which P10 deployed a persona pattern to review a controlled experiment, immediately followed by the description of the experimental design. In this case, P10 favored a minimal interaction over an iterative prompting approach.}
        \label{fig:other_patterns}
\end{figure*}

Contrary to single prompts that were deployed in early design phases, iterative prompting strategies were considered beneficial in later phases.
They often acted as mitigation when participants failed to generate a single prompt because they found it difficult to decide which information about the design task was most relevant to be provided as input to the AI to produce a reasonable outcome. 
In such cases, participants P6, P7, P9 and P10 deployed a flipped interaction pattern that flips the interaction so that the AI asks the user questions~\cite{white_arxiv2023}. Participants indicated that this helped to identify which information the Generative AI requires for being able to generate, for example, a reasonable experimental design from scratch. An example prompt that deploys a flipped interaction pattern can be found in Figure~\ref{fig:other_patterns}A. For most design activities, the AI asked relevant questions to collect the necessary input, leading to successful results. However, in one instance, when P7 worked on a technical design activity, the AI asked overly specific technical questions such as \aireply{What is the maximum energy density that the power source can have?}. 
These questions were not compatible with the designer's mental model, surpassing her knowledge, and ultimately limited the practical usefulness of the flipped interaction approach for this rather technical activity.

For rather technical design activities, including component selection (P7) and haptic experience design (P5), iterative prompting strategies that gradually dive deeper into the design activity together with or under the guidance of the designer have proven more useful. 
Hereby, instead of letting AI ask questions before generating a solution, the AI and designer jointly generate a solution step-by-step. This allows the designer to steer the conversation, iteratively refine, and flexibly intervene in cases of erroneous or too generic output by providing details about the design task only when considered necessary. 
We provide example excerpts of the component selection in Figure~\ref{fig:component_selection}; for more details refer to Appendix~\ref{app:component_selection}. In the provided example, P7 initiated the interaction with Generative AI through a prompt that comprised (1) the context of the project, (2) the purpose of the prompt (component selection), and (3) the mode of interaction (Generative AI provides choices that gradually become more detailed). 
After the AI had then suggested components, P7 provided different forms of answers: Either, P7  selected a component and expanded on detailed options (types, models) for this component (4) and inquired about pros and cons for each. Or in case the offered components were not satisfactory, she explored alternatives (5), asking AI to list further options beyond the provided ones. Further, once P7 had chosen a specific component, she constrained Generative AI to only select further components compatible with that one (6). 
Finally, P7 concluded the conversation by asking for improvements on the selection of components, for recommendations on their quantity, the estimated cost, and on additional components needed for the final assembly (7).
These different ways by which the designer can react and steer the conversation demonstrate the flexibility that this strategy offers. Rather than automating the design activity, AI augments the decision-making process, thereby enabling the designer to retain control.

\subsection{Minimal Interaction for Gathering Feedback}
The findings presented above have primarily focused on producing new artifacts for design activities, whether from scratch or as refined iterations of existing artifacts. 
However, participants have also used Generative AI to gather feedback on the quality of artifacts produced during design activities, such as gathering feedback on an experimental design. 
The prompts deployed for this purpose were comparable in detail and structure: They explicitly outlined the expected aspects of the artifact in a single prompt, and hence prioritize minimal interaction with Generative AI. This stands in contrast with the iterative strategy outlined above where the participants successively interacted with the AI to create a solution together. 
In addition, contrary to the structured prompts deployed when generating designs from scratch, the prompts used to gather feedback through Generative AI provided only minimal additional information about the design task and did not follow a very explicit structure. 
For instance, P10 deployed a prompt that comprised only three key elements in a single prompt (see Figure~\ref{fig:other_patterns}B): (1) introduction of a persona pattern (UX researcher), (2) the purpose (review experiment design, identify issues and concerns), and (3) the experimental design (including the variables, sample size, target population, and study procedure). This was a successful strategy to uncover flaws in the study design. We attribute the effectiveness of minimal interactions to the more open-ended nature of the designer's request when gathering feedback, aiming to uncover overlooked issues or new insights for the provided artifact. This stands a contrast to using highly structured or iterative prompts, which allow the designer to steer the output more purposefully in desired directions.


\subsection{Adaptive Use of Persona Pattern}

Finally, our analysis revealed that persona patterns were deployed throughout various design phases, but with different purposes and adapted to the respective phase. 
Persona patterns give the AI a persona that helps it select what types of output to be generated~\cite{white_arxiv2023}.
P3 and P4, who worked on conceptual design activities deployed persona patterns primarily to guide the AI to take on the varied perspectives of different stakeholders, in order to better understand potential critiques and points of improvement on the ideas. 
For instance, P3 generated design alternatives from a specific stakeholder's perspective by prompting: \pquote{Assume you are a 22-year-old student studying Bachelor in Game Design who is a pro-gamer and interested in VR gaming}. To receive specific insights into how to improve a conceptual design for a specific user group, he prompted: \pquote{Assume you are a teen boy with left leg disability.}

This use of the persona pattern stands a contrast to P9 and P10, whose goal was to identify a suitable evaluation method. 
For this purpose, they deployed the persona pattern of a UX designer and prompted: \pquote{You are an experienced UX researcher} (P10) or \pquote{You are an evaluation expert and would help the user identify the right method (high-level) for the evaluation given specific criteria that need to be evaluated} (P9). This type of persona helps to guide the AI toward behaving like an expert, a valuable strategy when a designer seeks targeted advice.

\section{Quality of AI-Generated Artifacts}
We now analyze the strengths and limitations of the AI-generated artifacts presented as well as more principled challenges our participants faced across design activities and phases.

\subsection{Strengths and Limitations of the Generated Artifacts}
\begin{figure*}
        \centering
        \includegraphics[width=\textwidth]{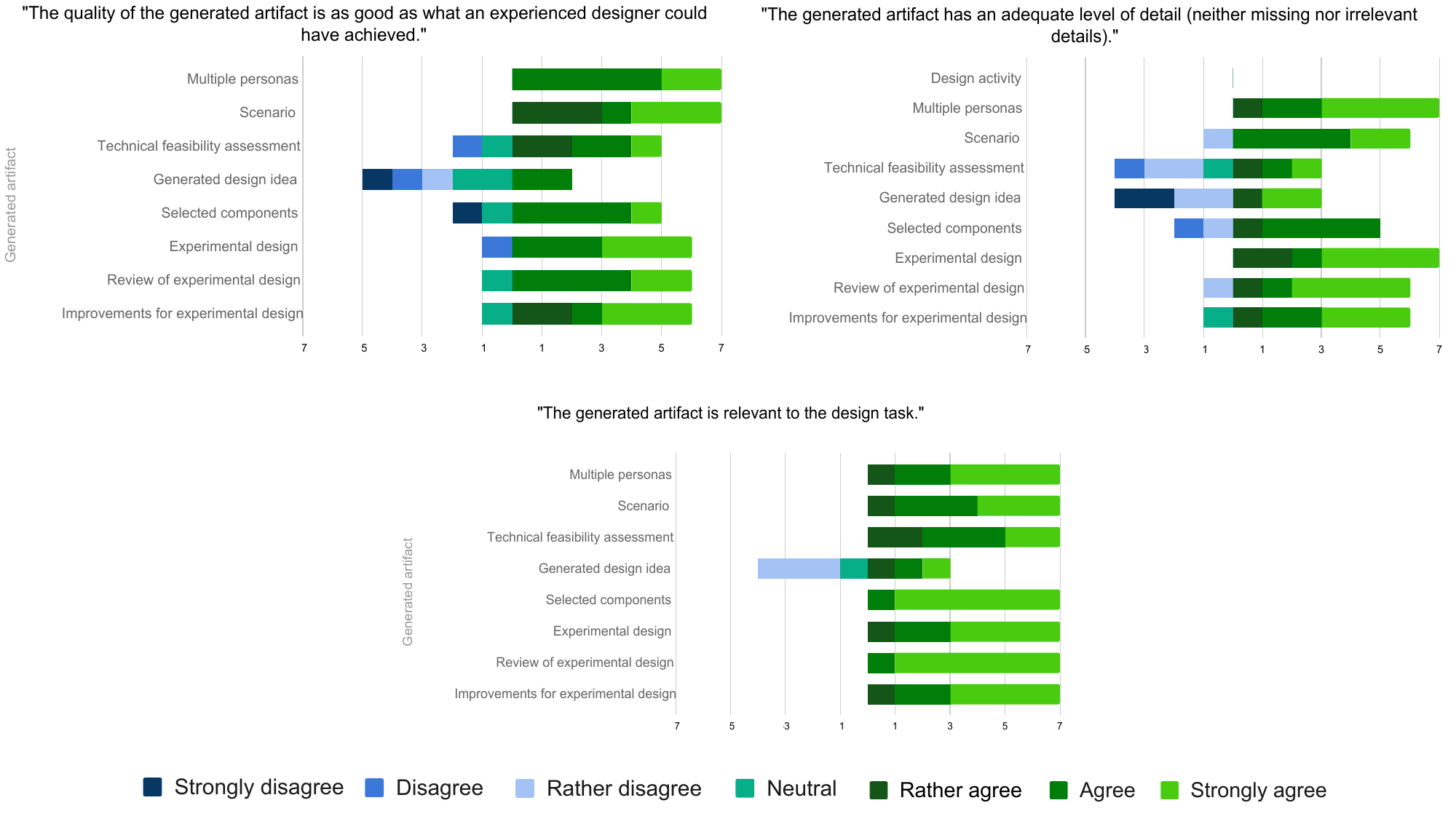}
        \caption{Answers to the three items in the Likert questionnaire gathered in the second user study. The items evaluate the quality, level of detail and relevance of the generated artifacts. The x-axis displays the number of participants reporting positive responses (right of the central diverging point) versus neutral to negative responses (left).
        Overall, participants acknowledged the relevance of the artifacts to the provided design task, but were more critical of the quality and the adequacy of the provided level of detail. 
        }
        \label{fig:quantitative}
\end{figure*}
In the following, we present the quantitative and qualitative results of the second user study, structured by the key phases of user-centered design. Figure~\ref{fig:quantitative} shows the participants’ answers to the three items in the Likert questionnaire. 

\paragraph{Requirements analysis}

The personas generated in the requirements analysis phase (presented in Appendix~\ref{app:persona_generation}) adequately represent potential stakeholders with different backgrounds who each have a relevant relation to VR gaming. 
The personas further cover essential aspects of those stakeholders, for instance, different pain points and challenges, such as the graphic designer who favors aesthetics, the software engineer who favors preciseness and reliability, or the high school student who wants a comfortable and stylish device. 
These successfully point to relevant requirements for the novel device with an adequate level of detail.
This was confirmed by the study participants who reported a high median of $\bar{x} = 6$ for the quality of the generated personas, and a very high median of $\bar{x} = 7$ for the adequacy of the provided detail and relevance of the artifacts for requirements analysis. 
Our study participants described the personas positively as \pquote{diverse} (P11), and noted that \pquote{especially the coherence of the character details is particularly good} (P13). 
Some aspects mentioned in the personas, however, do not appear human-made and stand out. P17 stated that \pquote{an experienced designer [\ldots] could have slightly better perceptions of pain points}. For instance, mentioning the \aireply{hygiene of the device, especially when sharing it with others} as one of the key pain points, or the desire to feel \aireply{more feedback on his body, such as direction, speed, and force} of a user appears highly specific and rather artificial.
This indicates the need for manual fine-tuning AI-generated personas to remove unrealistic character traits.

Similar to the generated personas, the AI-generated scenario (see Appendix~\ref{app:scenario}) successfully pinpoints different relevant design requirements, such as allowing the user to define the location where haptic feedback is provided, or to provide the option to calibrate its intensity. 
Overall, the scenario is detailed and walks the designer through important steps of the interaction with the device as well as the setting of the overall scenario. As such, it fulfills the principled requirements of a scenario and can be useful in supporting the design process. 
This was also pinpointed by our study participants who rated the quality, level of detail and relevance of the presented scenario with a high median of $\bar{x} = 6$. P11 stated that \pquote{The presented scenario is detailed and believable. [\ldots] I find that an experienced designer would describe a scenario similarly}.
Copilot further suggested variations of the provided scenario. 
These, however, put a focus on details, which are not relevant to the actual interaction with the device, such as details about the mode of the VR game. This was also pointed out by some participants who, for instance, \pquote{would expect a bit more focus on haptics} instead~(P12). Therefore, the generated scenario can provide only limited guidance on alternative outcomes and further requirements of the device.

\paragraph{Conceptual and physical design}
P4 used Copilot to assess the technical feasibility of potential designs (see Appendix~\ref{app:evaluate_concept}). 
The results demonstrate that the AI can indeed point to major technical challenges. Hence, our study participants reported the quality and relevance of the generated artifacts rather positively, with a median of $\bar{x} = 5$ and $\bar{x} = 6$, respectively. P12 stated that \pquote{for beginners, this would be a good entry point}. 
However, it became apparent that the recommendations were overly geared toward the positive: \pquote{No matter how ambitious the idea, the conclusion of the report seemed highly positive, that the idea is feasible, with the recommendation to pursue the idea}~(P4). 
Since design typically involves generating many ideas and then selecting only few of them for further investigation, it is as important to identify which ideas are \textit{not} worth pursuing further as it is to identify strong ideas. Therefore, a too positive stance about all ideas can be misleading, particularly for inexperienced designers who seek advice from Generative AI.
Furthermore, the generated artifacts are rather brief and lack more detail that would allow for a more nuanced understanding of the technical feasibility, beyond the big challenges. Hence they were sometimes described as \pquote{vague} (P14, P15), assuming that \pquote{an experienced designer would add more details}~(P14). This is reflected in participants ratings, who reported the level of detail with a median of only $\bar{x} = 4$. 
Consequently, while helpful for inexperienced designers, these aspects might offer less help for experienced designers.

When being prompted to iterate on existing designs to generate an idea for a novel haptic device (see Appendix~\ref{app:modify_hardware}), the resulting outputs displayed varying degrees of quality. Visually, the generated designs were appealing. Participants described them as \pquote{creative} (P13) and \pquote{futuristic} (P12), but also as \pquote{complex} (P12) and \pquote{packed with information and thus a bit overwhelming instead of helpful for the design process} (P15). Hence, they reported the adequacy of the level of detail provided in the artifact with a rather low median of $\bar{x} = 3$. Notably, one of the generated options, a suit integrated with robotic arms, appears innovative and aligns well with the design specifications, suggesting a promising conceptual direction. However, the two other generated ideas did not meet the design specifications of a robotic device and full-body haptic feedback, which was also noted by several participants (P11, P13, P14, P16). Particularly the proposed concept of earbuds with haptic features not only deviated from the prescribed form factor, but also lacks practical feasibility. Consequently, participants reported a median of $\bar{x} = 4$ for quality and relevance to the design task.

Finally, when using Copilot to choose components for the selected device design (see Appendix~\ref{app:component_selection}), the AI provides valuable guidance to identify key components, including sensors and actuators, but also secondary components, like wires, textile and zipper. 
In addition, the suggested quantities of sensors (14) and actuators (40) needed for the design appear reasonable, as they are within a range of what is suggested in most recent commercially available technologies, such as for instance the TactSuit\footnote{TactSuit: \url{https://www.bhaptics.com/tactsuit/tactglove}, last accessed 04/30/2024} that integrates between 16 to 40 vibrotactile actuators  in the vest, or the Teslasuit\footnote{Teslasuit: \url{https://teslasuit.io/products/teslasuit-4/}, last accessed 04/30/2024} that comprises 14 IMU sensors and 80 channels for haptic feedback. 
However, in the provided artifact, the AI states that only one microcontroller is required to control both the selected actuators and sensors, but the provided microcontroller does not offer enough pins to accommodate the indicated quantities of actuators and sensors. Similarly, P13 noted: \pquote{You cannot run the device for more than a few minutes [with one battery]. And 1 HUGE battery contradicts the requirements. Thus, I think this is wrong information}. 
Hence, human technical expertise is required for sanity checks to ensure that the identified components can be assembled accordingly. 
Despite these inaccuracies in the artifact, participants overall reported a median of $\bar{x} = 6$ for the quality and level of detail of the suggestions and a very high median of $\bar{x} = 7$ for their relevance to the design task, as it would provide designers with a \pquote{valuable} (P11) and \pquote{helpful overview} (P17). P12 summarized that \pquote{it is a good start for beginners for a further research on components}.

\paragraph{Evaluation}
In the example provided to generate a controlled experiment design (see Appendix~\ref{app:experiment_design}), Copilot suggested an experimental procedure that covers the main steps of how to conduct the experiment. It therefore provides designers with a good scaffold for an experimental procedure, which they can fill with more details, for instance about the exact task that study participants will be asked to carry out. 
In addition, the AI identified dependent, independent, as well as confounding variables. 
However, few participants identified some problems with the variables and coherence in the study design. 
P12 noted an inconsistency between the suggested procedure and the independent variables, as 'wearing the device' or 'not wearing the device' was not listed as an independent variable but presented as such in the generated procedure description. Similarly, P13 stated: \pquote{The AI says that the time played is equal for every participant (see confounding [variable]), but we measure the completion time for each participant? Either this is a contradiction or there is not enough detail on the distinction between completion time and play time}. 
Furthermore, the indicated variables, such as the dependent variable comprising \aireply{a questionnaire that assesses the level of immersion}, require further refinement as it \pquote{is missing detail} (P16). 
While these details require further iterations and careful manual validation, participants agreed that it provides \pquote{overall [a] good basic structure for [the] study design} (P16), which will \pquote{help answering the RQ} (P12).
Despite having received some criticism, the overall quantitative responses of the study participants were still positive as a result, with medians of $\bar{x} = 6$ for quality and $\bar{x} = 7$ for both relevance and an appropriate level of detail.

Furthermore, when provided with a flawed experimental design (see Appendix~\ref{app:evaluate_experiment}), Copilot successfully identified major flaws in the study and suggested further improvements. Participants were positive about this: \pquote{It did not only find the missing counterbalancing fact, but also provided other (in my opinion) useful tips (e.g., the standardized questionnaires)} (P13). They reported on a high median of $\bar{x} = 6$ for the quality of the generated artifact, and a very high median of $\bar{x} = 7$ for both, detail and relevance to the provided information of the designer. 

Finally, P10 evaluated whether Copilot can improve the study design based on insights gained in a pilot study (see Appendix~\ref{app:modify_experiment}). While most recommendations on how to improve the study were correct and useful, e.g. the suggestion to add a training session, some participants found them slightly too generic: \pquote{I believe a more thorough response about all--when the designer themselves seems inexperienced--would’ve been better--but I guess with [the] next prompt it’ll be addressed} (P7). 
Overall, participants were positive about the provided recommendations and reported the quality and detail of the generated artifact with a median of $\bar{x} = 6$ both, and the relevance with a median of $\bar{x} = 7$. 
However, the generated suggestions further require a critical sanity check. For instance, Copilot suggested increasing the number of trials to address learning effects. As noted by P3 only, increasing the number of trials is not useful in avoiding undesirable learning effects.
While experienced designers might be able to identify this as an incorrect advice, inexperienced designers can be misguided by this suggestion.

Next, we delve into principled challenges our participants faced across the specific design activities.

\subsection{Principled Challenges in Generating Artifacts}\label{subsec:challenges}
Some problems our participants encountered relate to principled shortcomings of common Generative AI tools, such as hallucinations~\cite{nah_routledge2023}, lack of depth and failure to understand the nuances of a design challenge~\cite{tholander_dis23,arora_arXiv23} linked to difficulties in providing an appropriate level of contextual information to avoid overly generic outcomes~\cite{skjuve_cui2023}, as well as biased outputs~\cite{nah_routledge2023,arora_arXiv23}. 
This poses a principled risk of overreliance~\cite{nah_routledge2023}, which is particularly critical when designing for and with human subjects. In the following, we detail on challenges specific for user-centered design:

In the early phases of the design cycle, our participants observed a convergence to stereotypical representation of users and scenarios, which they tried to minimize through structured prompts and the usage of persona patterns. 
For instance, P1 and P2 observed a tendency to converge to stereotypes during requirement elicitation. Throughout the study, P1 struggled with creating a diverse set of personas as \pquote{all manners of persona-generations ended up as a 26-28 year old software developer/designer named Alex who either worked on or was very interested in video games}. Similarly, P2 stated that the generated scenarios oftentimes cover VR games either situated in a medieval- or a space-setting.
Participants P3--P5 who worked on conceptual design further noted that the generated designs for wearable devices tend to cater to majority bodies unless explicitly prompted otherwise. 
This is critical as such stereotypical assumptions can harm the realistic assessment of requirements and ultimately the system design. A continuous human quality assessment is essential to detect and counter this.

\begin{figure}
    \centering
    \includegraphics[width=0.55\textwidth]{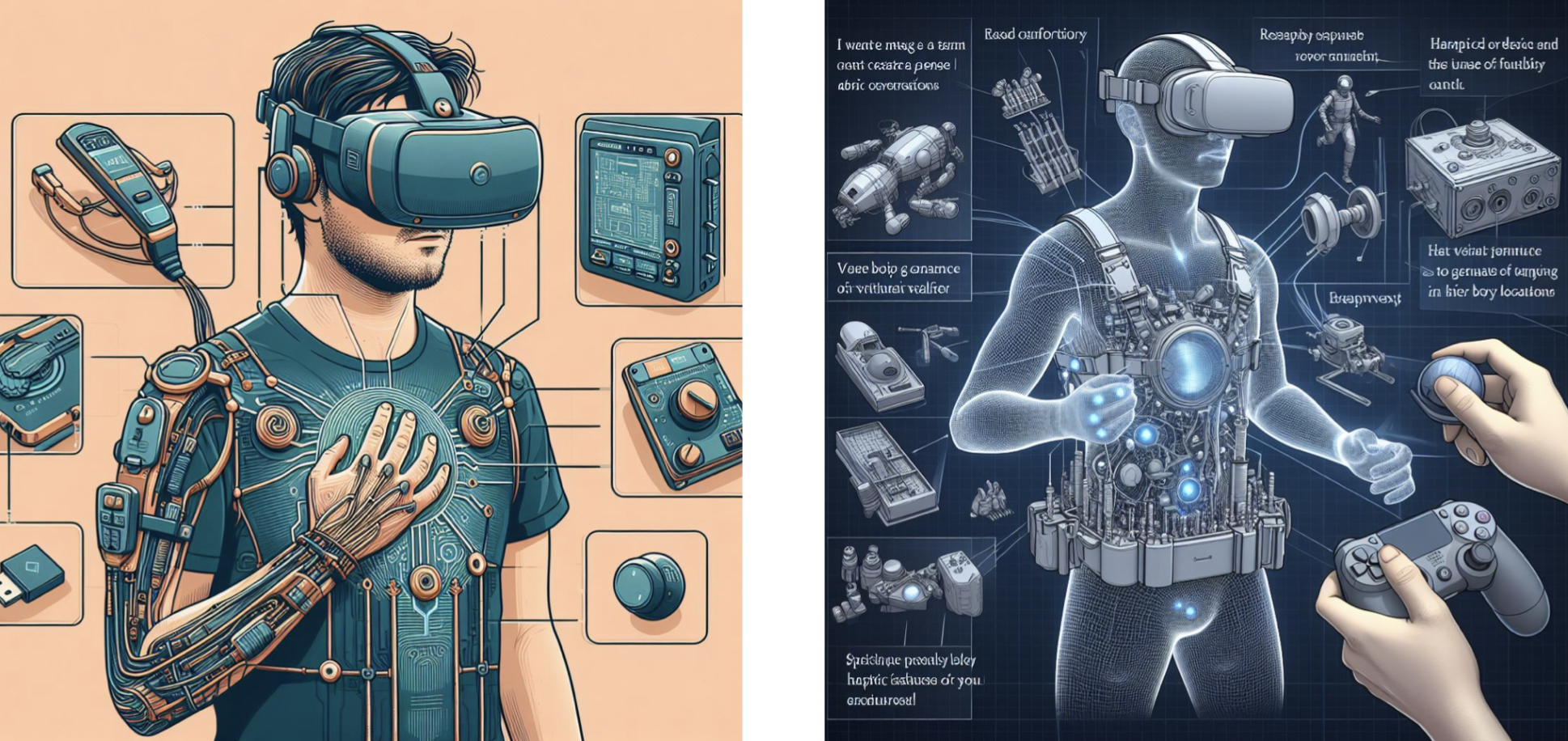}
    \caption{Two designs for a haptic feedback device generated by P7. The Generative AI was prompted to focus on the electronic components needed to build the device. While the generated images are visually appealing, they fail to depict realistic components.}
    \label{fig:hardware_problem}
\end{figure}

Contrary to the activities in the earlier design phases, participants questioned the usefulness of the artifacts in the later design activities where Generative AI was used to generate or iterate on functional designs.  
This relates particularly to tasks involving image generation where DALL-E prioritized visual appeal over functionality. 
Participants P6--P8 clarified: \pquote{For instance, when designers aim to simulate various combinations of materials or components, the generated outputs often exhibit a semblance of reality but fall short of aligning with existing objects, let alone accurately representing their assembly and connections}. 
Figure~\ref{fig:hardware_problem} depicts an example in which DALL-E failed to depict realistic electronic components although prompted so. 
This problem was aggravated as there were no possibilities to edit or locally manipulate the generated artifact. 
Participants P6--P8 further highlighted that this limited control over generated designs poses an essential challenge for both conceptual and physical design activities, because \pquote{in [the physical design phase], we’re looking for practical results like codes or solid prototypes. 
Naturally, this leads to the need to have some control or make edits to the final outputs. However, regular chat-based Generative AI tools have limited control options for the final results due to their restricted interaction choices, mostly using text or image prompts}. 
It will be imperative to develop interfaces that support interactions which align with the iterative nature of the design activities, for instance by offering more direct ways of manipulating generated output.

Participants who worked in the earlier design phases (P1--P6) further encountered problems in conveying and generating descriptions of haptic experiences. 
While it is already challenging to express haptic experiences in a non-ambiguous way in human-human interaction, participants reported on fundamental problems with Generative AI's inability to distinguish between haptic experiences and human feelings. 
This was evident, for example, in generated outputs stating \aireply{a haptic device that can provide intense and variable feedback on different parts of the body, such as the feeling of pain from a wound or the thrill of a jackpot} (P1) or a haptic device that \aireply{can simulate various sensations such as [\ldots] emotions} (P6). 
From the designer's point of view, this limits the usefulness of AI support in the design of haptic experiences.

Finally, the participants have raised ethical concerns of Generative AI. 
Participants' concerns encompassed unregulated data management (P1, P2, P6--P8) and intellectual property concerns in the conceptual and physical design phases, not knowing whether and how much the generated artifacts resemble existing products (P3--P8). These principled ethical concerns are subject to discussion in an increasing body of work~\cite{nah_routledge2023,arora_arXiv23,rezwana_cc2023}.
One major practical problem we encountered in our study revolves around a few major companies controlling access to Generative AI tools, with sometimes intransparent policies. 
Initially, we had started our exploration with another company's tool. However, the study accounts were deactivated without a comprehensible justification, implying that all our participants' data of previously used prompts was lost.
Until the publication time of this article, the company has not responded to our request for explanation and for providing us with our data.  
The lack of transparency in access restriction and data management, although within terms of service, raises ethical concerns which extend beyond interaction design.

\section{Implications and Discussion}
Based on our findings, we derive implications for designers interested in using Generative AI and implications for guiding the future development of Generative AI tools for interaction design. We conclude with a general discussion on the role of Generative AI in interaction design and outline areas for future work.

\subsection{Implications for Designers}
Our results demonstrate versatile strategies of how Generative AI can support designers in the user-centered design cycle. For instance, designers can leverage existing Generative AI tools to create a set of distinct personas and detailed scenarios, rapidly draft experimental designs, or adopt co-creative approaches where the designer and AI jointly select components for the intended prototype design. While many examples leverage AI to generate artifacts from scratch or iterate on existing ones, our study also demonstrates the usefulness of Generative AI to assess the quality of and provide feedback for existing artifacts of design activities, such as identifying major flaws in a given experimental design.

Furthermore, our results reveal that the strategies and prompt patterns should be adapted to the design phase and specific design activity in order to be successful. 
To create solutions in the early design phases, single prompts that provide a structure for the output were particularly beneficial.
They can ensure a certain level of detail and diversity of generated designs, which particularly eases the exploration of alternative outcomes for design activities. The specific structure of the prompt, however, varies depending on the design activity. We recommend meta-prompting as a useful strategy for creating these structured prompts. 
To create solutions in advanced design phases, such as physical and experimental design, iterative prompting strategies are advisable. Here, our participants either deployed a flipped interaction pattern or guided the AI to gradually dive deeper into the design activity.
Contrary, when leveraging AI to assess the quality of an artifact of a design activity, less structured single prompts resulted in satisfactory outcomes.
Finally, throughout the design phases, persona patterns can help the designer to steer the AI to take on varying perspectives of stakeholders, whereas the persona pattern of a UX designer can be deployed to guide the AI toward behaving like an expert who provides targeted guidance and advice.

While Generative AI can offer versatile support, speed up the early iterations of the design process, and provide guidance for inexperienced designers, we urge designers to critically evaluate the quality of the generated artifact. Designers should be mindful of biases, such as stereotypical representations of stakeholders, or misinformation that can affect all design phases. Furthermore, although the successful prompting strategies were the result of extensive iterations by our study participants, the generated artifacts presented in this article require further refinement to come close to the quality of a final solution. This can be achieved either by manually refining the artifact or with the assistance of AI. Finally, we emphasize questioning and clarifying ethical concerns, such as intellectual property, when leveraging Generative AI for the conceptual and physical design phases.

\begin{figure*}
        \centering
        \includegraphics[width=\textwidth]{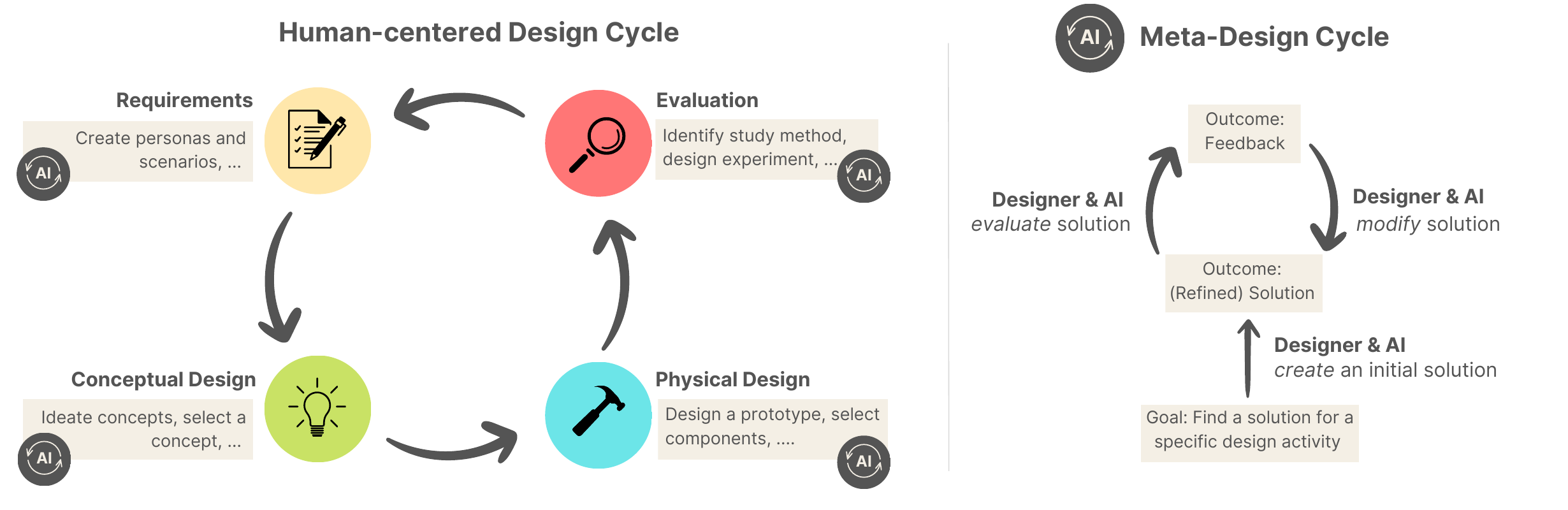}
        \caption{The user-centered design cycle comprises requirement elicitation, conceptual and physical design, and evaluation. We propose an iterative AI-in-the-loop meta-design cycle in which designer and Generative AI jointly create, evaluate, and modify a solution for the individual design activities of each design phase.}
        \label{fig:design_cycle}
\end{figure*}

\subsection{Implications for Future Tools}
Our study revealed several challenges that inform the design of future Generative AI tools.
Firstly, extending on insights from previous work~\cite{dang_arxiv2022}, Generative AI tools that leverage text as input to generate artifacts for diverse design activities should assist designers in formulating more effective prompts. This involves helping designers understand the context in which the AI operates and how the provided information influences the generated artifacts. This further involves improvements in resolving or flagging ambiguities when designers describe subjective experiences, such as haptic experiences -- an area where current models fail~\cite{liu_arxiv2023}. 

Furthermore, domain-specific tools, datasets, or plug-ins would help to combat biases in generated designs. For instance, a dataset targeted at persona generation that aims to cover diverse user groups will likely reduce the reproduction of stereotypes. 
GPTs, which allow encoding extra knowledge to customize ChatGPT, are a first promising step in this direction.
Additionally, providing designers with tools or plugins to actively identify, flag, and address biases in real-time, another emerging research area~\cite{bai_arxiv2023}, can empower them to create more inclusive and diverse designs.

Considering the importance of explainable AI, we further envision future tools that communicate transparently how the generated artifacts of conceptual or physical design phases were influenced by or resemble existing designs in the dataset. This could be realized through similarity metrics or by enabling the designer to actively explore the data in the underlying dataset.

Tools and plug-ins that leverage image generation models for conceptual and physical designs should prioritize functional elements, such as electronic components, over visual fidelity, allowing designers to focus on the practical aspects of their designs. 
Moreover, enabling designers to locally manipulate these generated designs, as suggested in DirectGPT~\cite{masson_arxiv2023}, is a promising extension that allows for effective post-editing.

Finally, a notable challenge lies in the absence of clear standards or guidelines to assess the quality of generated output. 
Establishing these standards and guidelines is essential for enabling designers to leverage Generative AI tools effectively. 
By providing designers with insights into the strengths and limitations of Generative AI tools, developers can empower them to make informed decisions and optimize their use of these tools in their design workflows. 
Yet, how to appropriately manage expectations is an open question~\cite{zamfirescu_chi2023}.

\subsection{On the Role of Generative AI in Interaction Design}
Our study highlights the pivotal role of Generative AI in supporting designers across various design activities and phases. The opportunity to use Generative AI, here LLMs, for a variety of design activities, and to do so through natural language, aligns with two fundamental design principles for co-creative systems~\cite{resnick_report2005}, emphasizing \textit{wide walls} and a \textit{low threshold}.
While Generative AI facilitates rapid design generation, exploration, and iterations, the generated artifacts necessitate manual quality assessments and human guidance to grasp the nuances of design. This could be seen as a principled limitation of Generative AI tools for general use, which provide broad but not necessarily deep domain-specific knowledge and hence do not fulfill the \textit{high ceiling} design principle of co-creative systems.
Extending on discussions in recent work~\cite{schmidt_interactions24,tholander_dis23}, we impose that the quality of general use Generative AI should not compete with expectations to perfectly execute user commands or automatically generate polished designs without the human-in-the-loop. Instead, we should interpret the potential of Generative AI in interaction design as a unique chance to complement the designer's expertise.
Our findings allow designers to integrate Generative AI in an iterative process interleaved with manual quality assessment and refinements. In this process, the designer can (1) create an initial draft for a specific design activity with the help of Generative AI, (2) evaluate it with Generative AI and using designer's own judgments, and (3) modify the solution together with Generative AI based on the provided feedback. 
This process can be iteratively continued starting at step (2), until converging to a solution for the specific design activity that fulfills the designer's needs. This new process can be considered a meta-design cycle that is applied for an individual design activity within any phase of the user-centered design cycle. This is illustrated in Figure~\ref{fig:design_cycle}. 
We consider it an important step towards a future in which interaction design is the result of a collaboration between Generative AI and the designer.

\subsection{Limitations and Future Directions}
The participants of our first study were early adopters of Generative AI drawn from an academic context.  
This might have facilitated the entry in integrating Generative AI into the design activities, as participants might have already calibrated expectations of Generative AI's capabilities, established personal practices of prompting, and no deeply established personal workflows in the individual design activities yet. Future work could complement these insights with the perspectives of seasoned design practitioners.
The results of the second study may have been influenced by participants' limited use of Generative AI for the purpose of interaction design, potentially explaining the rather positive reactions and quantitative ratings. 

Furthermore, it is essential to clarify that the identified strategies and prompting patterns are not exhaustive, and we actively encourage fellow researchers to expand upon these.

Finally, recognizing the dynamic development of Generative AI, new architectures may bring new challenges. It remains to be seen to what extent specific prompts will generalize to future LLMs. Most certainly, specific details will change or may become obsolete in the future; for instance, tools might be able to automatically infer that a persona needs an engaging tone. Yet, we believe that the identified strategies and challenges themselves are of a more principled nature as they go beyond the level of specific prompt design. 
\section{Conclusion}
In this article, we reported on the results of a study which investigated the potential of Generative AI to support designers across the four key phases of user-centered design. 
Our results demonstrate that Generative AI can successfully support the designer in these key phases and that the generated artifacts require manual quality assessments and guidance to grasp the nuances of design.
We further provided detailed empirical accounts and practical insights of successful examples, emerging strategies, and recurring prompting patterns. The analysis of the results revealed that the practicability of identified prompting patterns varies across the design phases and activities. In early design phases which have limited information to convey, highly-detailed prompts were more successful in generating artifacts than in later design
phases, which benefited from iterative prompting patterns. To gather feedback for existing artifacts of design activities, less detailed single prompts proved useful. Persona patterns were deployed across the design phases, but with different purposes. We further identified additional challenges that our participants faced. These comprise convergence to stereotypical representation of users and scenarios in the early design phases, difficulties in generating functional designs and conveying technical or subjective information relevant to later design phases, alongside ethical concerns. From these findings, we derived implications for designers and future tools that support interaction design with Generative AI and propose a meta-design cycle to reflect the role of Generative AI tools in user-centered design. 
Ultimately, we hope our exploration provides practical and useful insights into emerging strategies and opens up promising directions toward effectively integrating Generative AI in interaction design.

\begin{acks}
We thank all participants of our user studies and acknowledge Artin Saberpour for his help in advising the groups. 
\end{acks}

\bibliographystyle{ACM-Reference-Format}
\bibliography{0-main}

\appendix
\clearpage
\section{Design Task Description}\label{sec:task_description}

\paragraph{Design Challenge} You have been asked to improve the gaming experience of VR applications, with a particular focus on the haptic experience.

\paragraph{Situation of Concern}
Haptic feedback can enable a more immersive VR experience by engaging our sense of touch. Existing solutions comprise, e.g., vibrotactile grids which are attached on the user’s skin and create haptic feedback through vibration or stationary robotic arms which are placed close to the user and create haptic feedback by touching the user’s body with their end effector. However, these approaches are either typically limited to creating haptic feedback only at one fixed body area (e.g., vibrotactile grids) or require the user to stay close to the device, limiting the user’s mobility (e.g., stationary robotic arms). There is a demand for a device that solves both of these challenges in order to improve the haptic experience.

\paragraph{Intended Solution}
The intended solution consists of a novel device for gamers that can create a sense of touch at different locations of the user’s body without restricting the user’s mobility. The device should be worn on the user’s body and be able to create a variety of haptic experiences at different body locations, such that it can be used for different VR gaming applications. As the device is worn on the body, it should be light-weight and easily attachable and detachable.

\newpage

\section{Creating Multiple Personas}\label{app:persona}

\subsection{An Unsuccessful Example of Generating Personas}\label{app:persona_generation_creative}
\begin{figure*}[h!]
    \centering
    \includegraphics[width=0.85\textwidth]{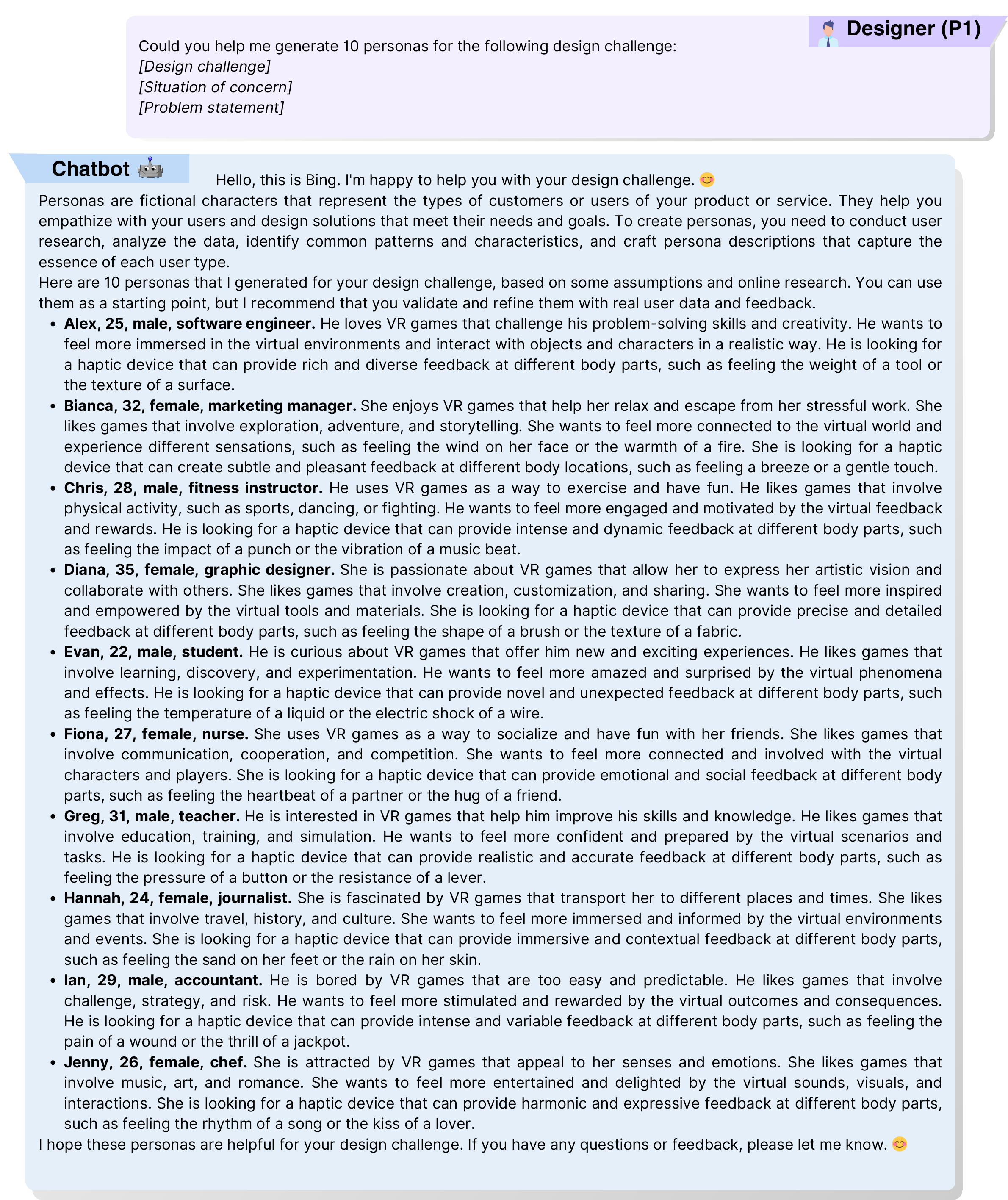}
    \caption{Detailed input and output developed by P1 to create personas. P1 used a single undetailed prompt. The generated personas lack detail.}
    \label{fig:persona_generation_creative}
\end{figure*}
\clearpage

\subsection{A Successful Example of Generating Personas}\label{app:persona_generation}
\begin{figure*}[h!]
    \centering
    \includegraphics[width=0.75\textwidth]{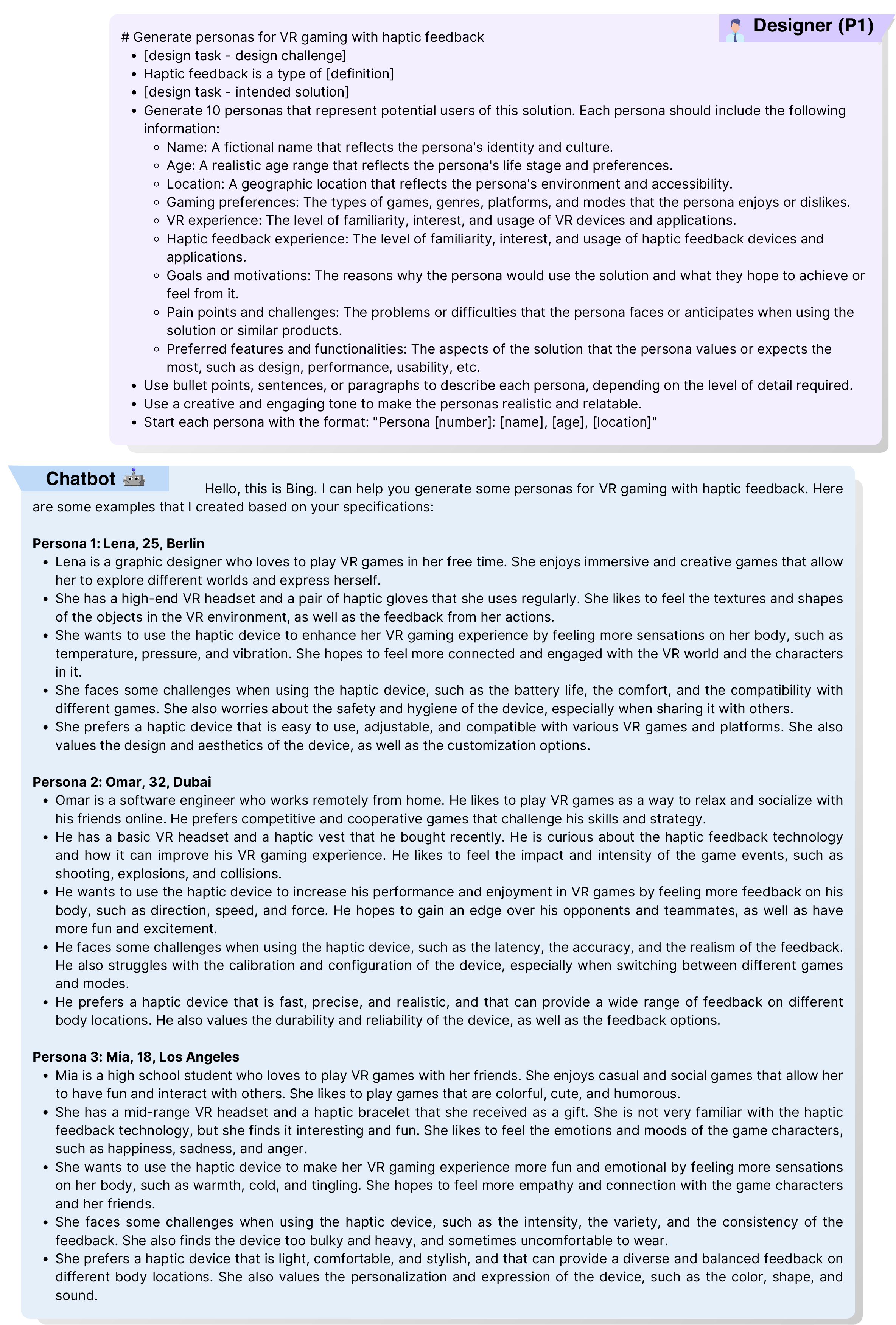}
    \caption{Detailed input and output developed by P1 to create personas. In this example, P1 used Copilot to mass-generate personas with a detailed prompt, resulting in more detailed personas.}
    \label{fig:persona_generation_new}
\end{figure*}

\clearpage
\section{Generating a Scenario}\label{app:scenario}

\begin{figure*}[h!]
    \centering
    \includegraphics[width=0.8\textwidth]{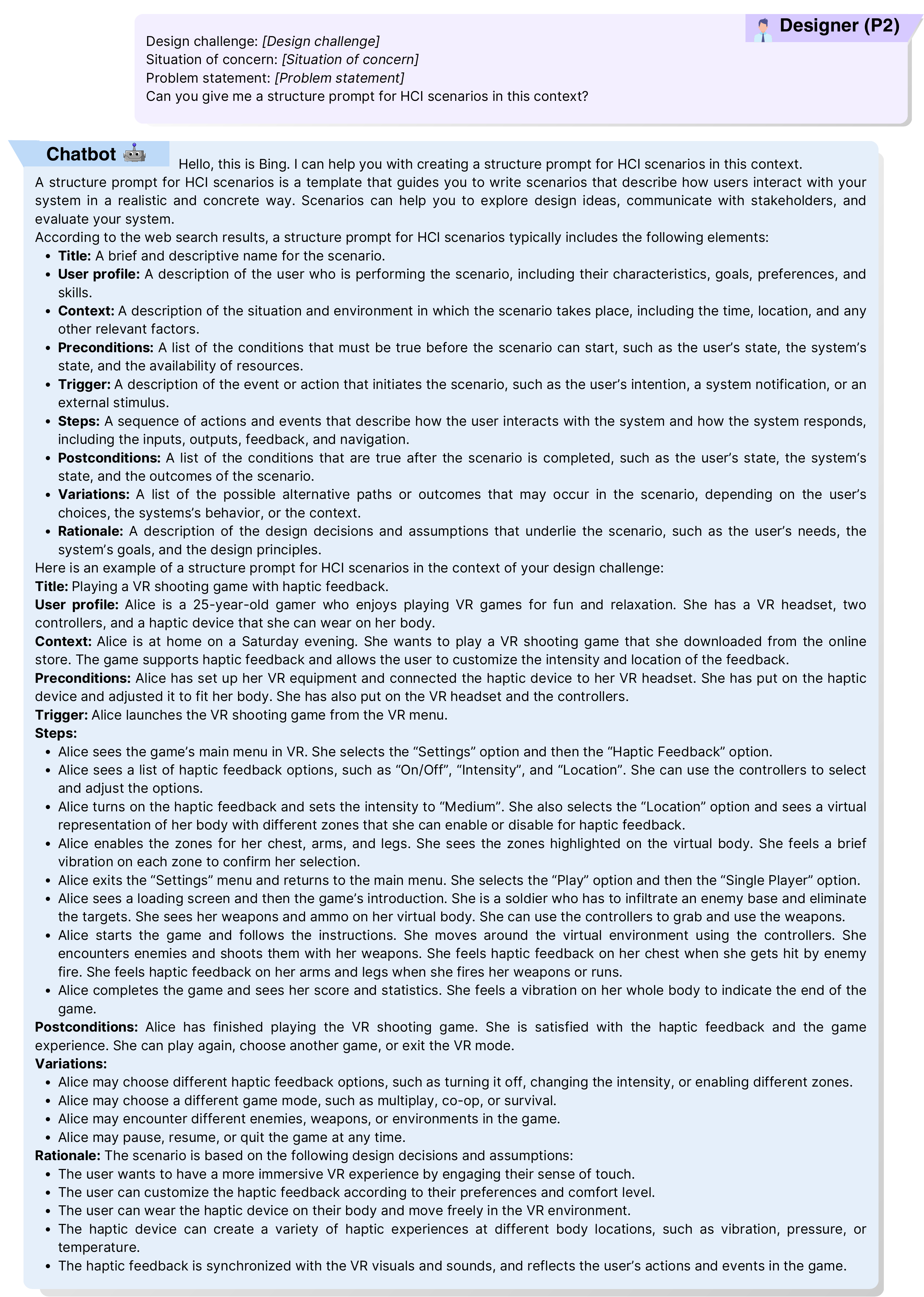}
    \caption{Detailed input and output developed by P2 to generate a scenario.}
    \label{fig:scenario_generation}
\end{figure*}

\clearpage
\section{Assessing the Potential of a Conceptual Design}\label{app:evaluate_concept}

\begin{figure*}[h!]
    \centering
    \includegraphics[width=0.9\textwidth]{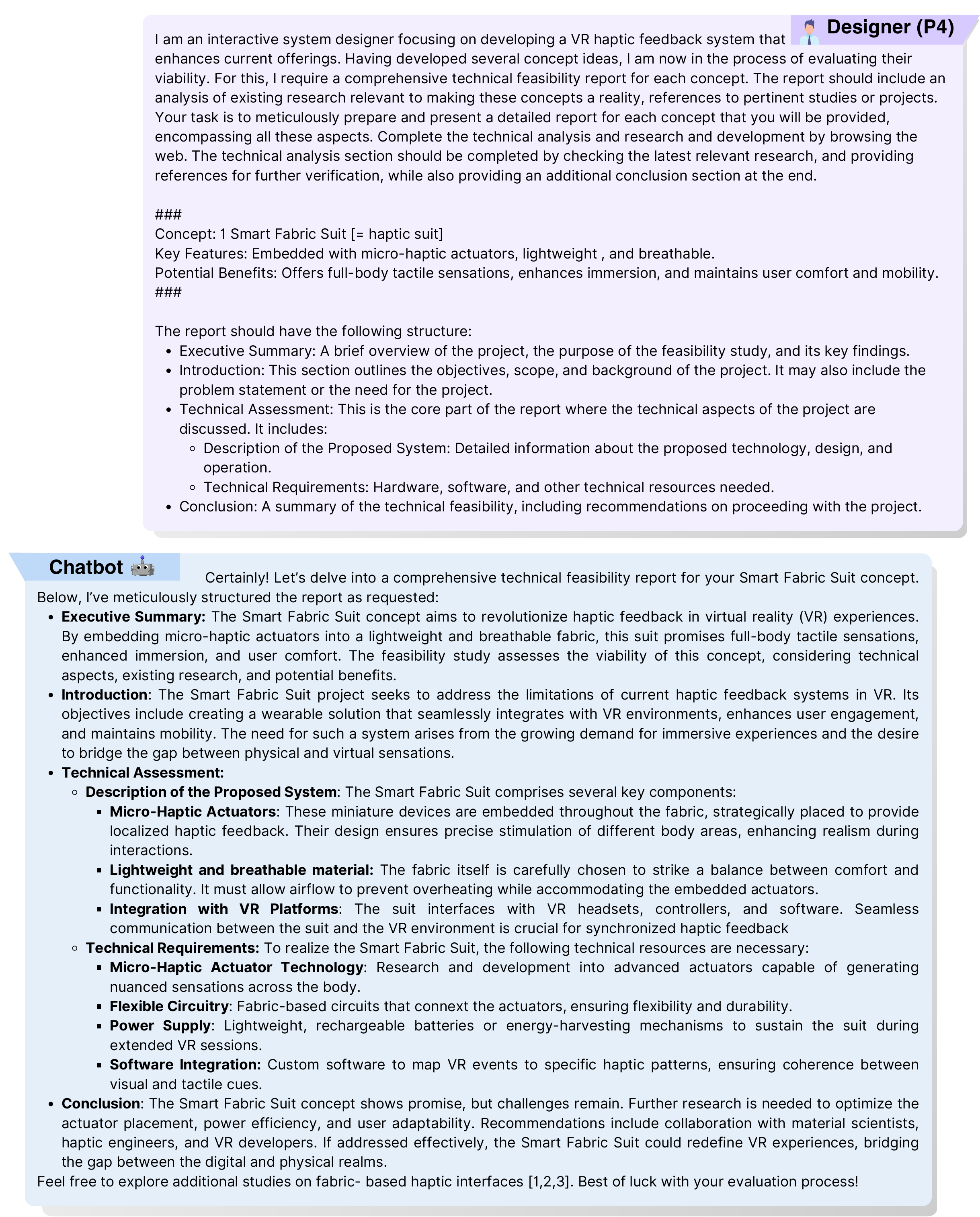}
    \caption{Detailed input and output developed by P4 to assess the technical feasibility of a concept using a detailed prompt.}
    \label{fig:evaluate_concept}
\end{figure*}

\clearpage
\section{Generating a Design Idea for a Physical Prototype}\label{app:modify_hardware}

\begin{figure*}[h!]
    \centering
    \includegraphics[width=0.8\textwidth]{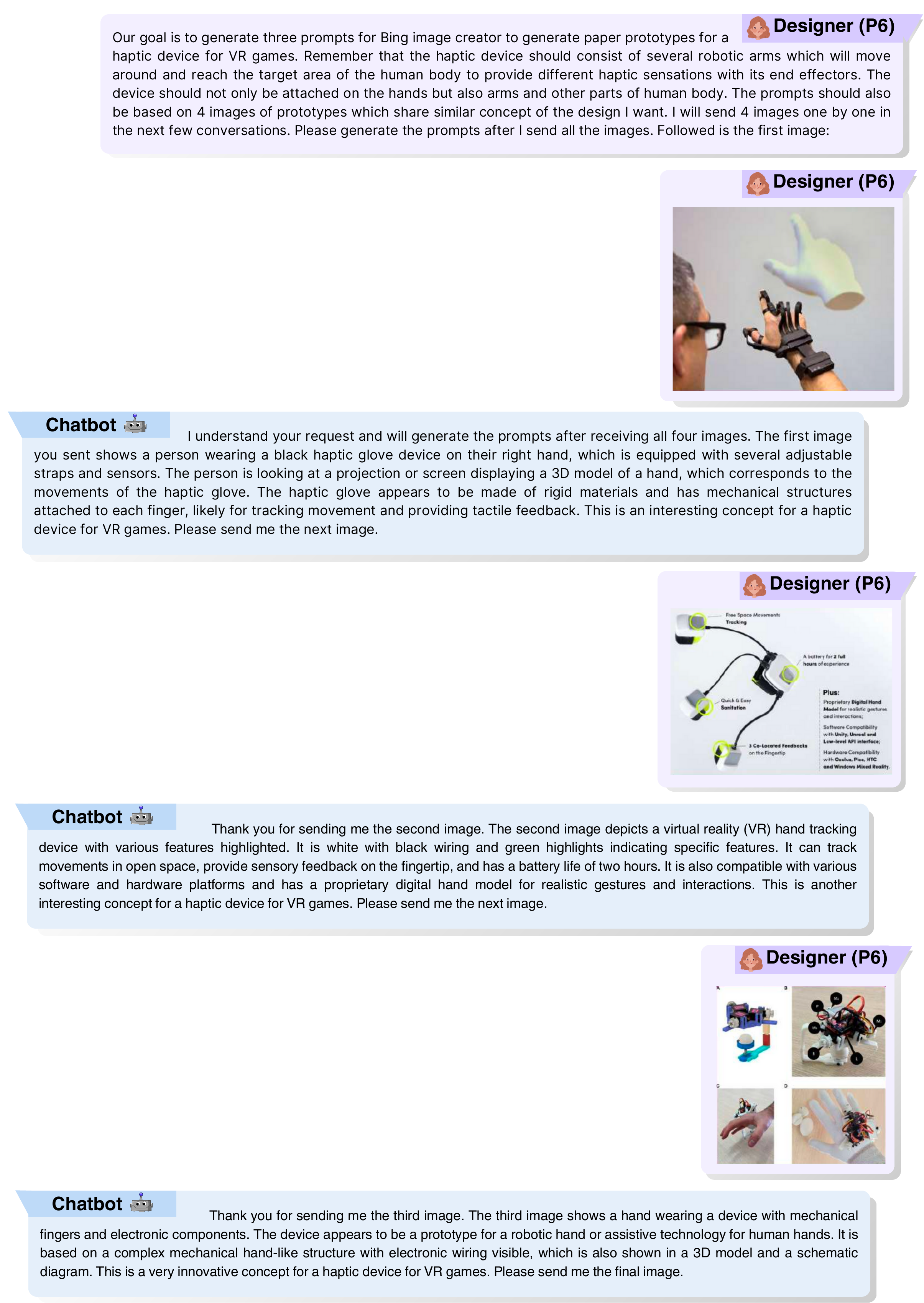}
    \caption{First part of the detailed input and output developed by P6 to iterate on the design of physical prototypes. P6 provided 4 photos of existing prototypes. Copilot automatically described the input images.}
    \label{fig:modify_hardware1}
\end{figure*}

\begin{figure*}[h!]
    \centering
    \includegraphics[width=0.9\textwidth]{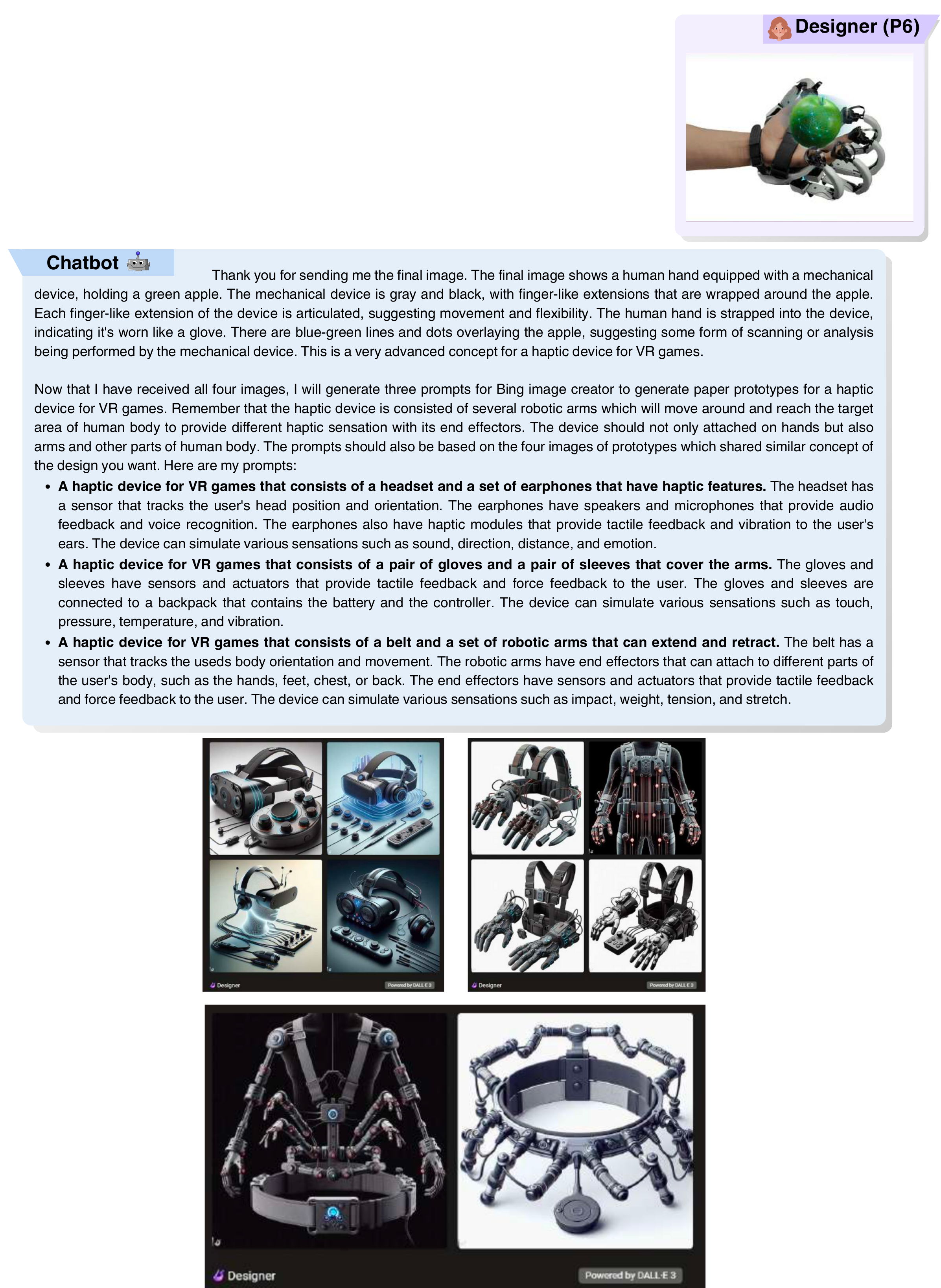}
    \caption{Second part of the input and output developed by P6. Copilot generated three image prompts. Based on these prompts, Bing Image Creator created several variations for each.}
    \label{fig:modify_hardware2}
\end{figure*}

\clearpage
\section{Selecting Components for a Hardware Prototype}\label{app:component_selection}

\begin{figure*}[h!]
    \centering
    \includegraphics[width=0.9\textwidth]{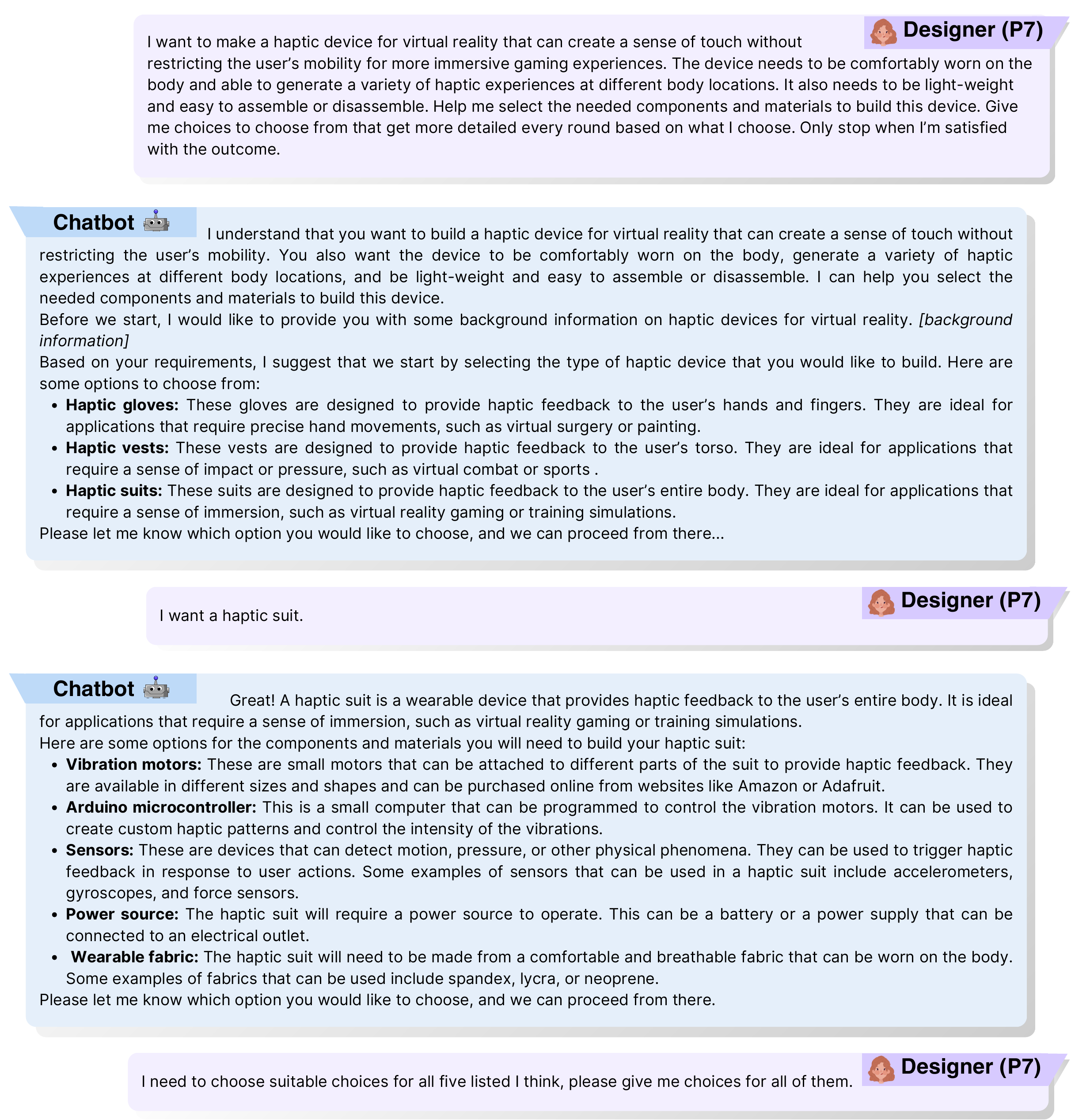}
    \caption{First part of the input and output developed by P7 to select components for a haptic suit. P7 co-created the list of components step-wise together with Copilot.}
    \label{fig:component_selection1}
\end{figure*}

\begin{figure*}[h!]
    \centering
    \includegraphics[width=0.9\textwidth]{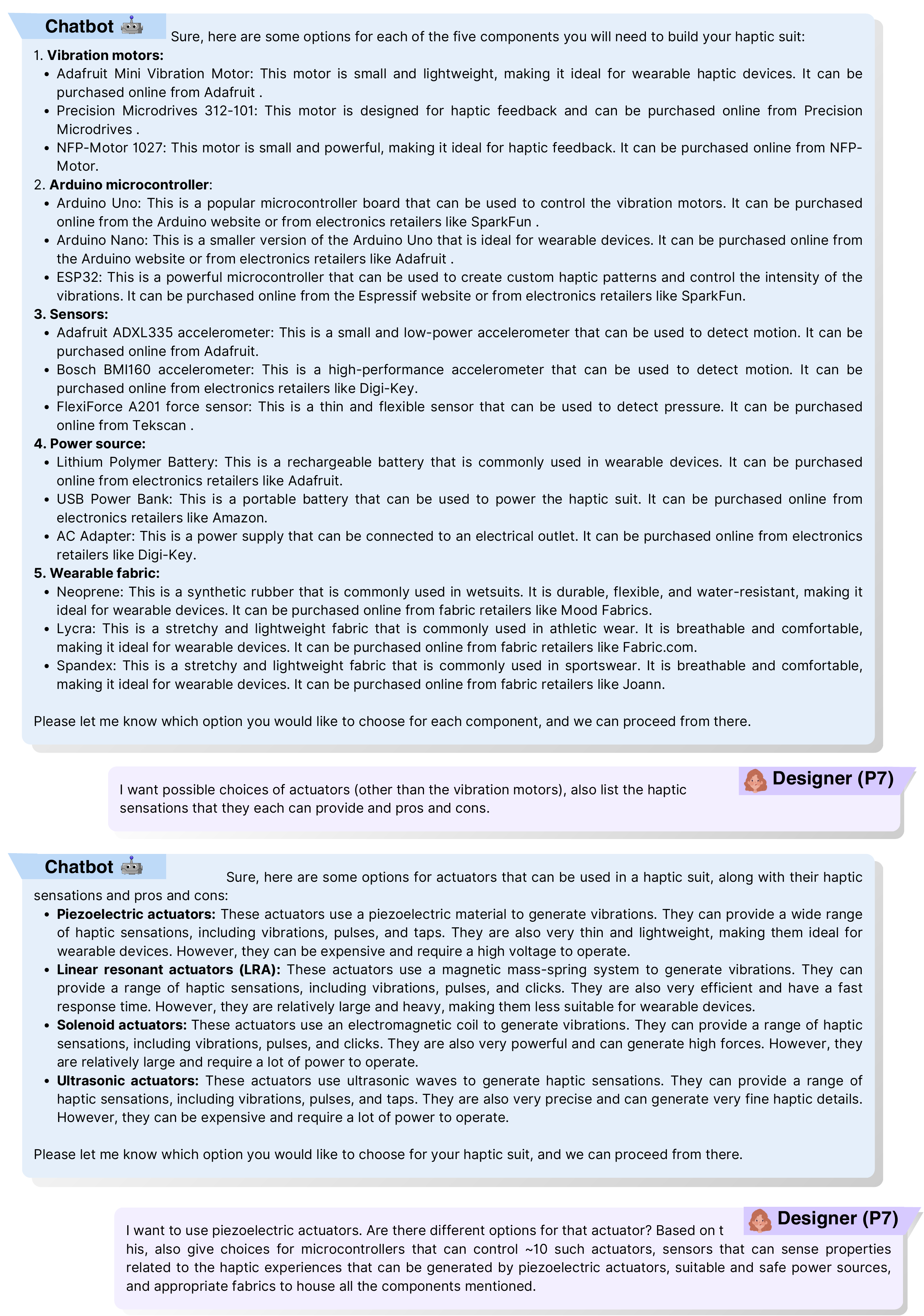}
    \caption{Second part of the input and output developed by P7 to select components for a haptic suit.}
    \label{fig:component_selection2}
\end{figure*}

\begin{figure*}[h!]
    \centering
    \includegraphics[width=0.9\textwidth]{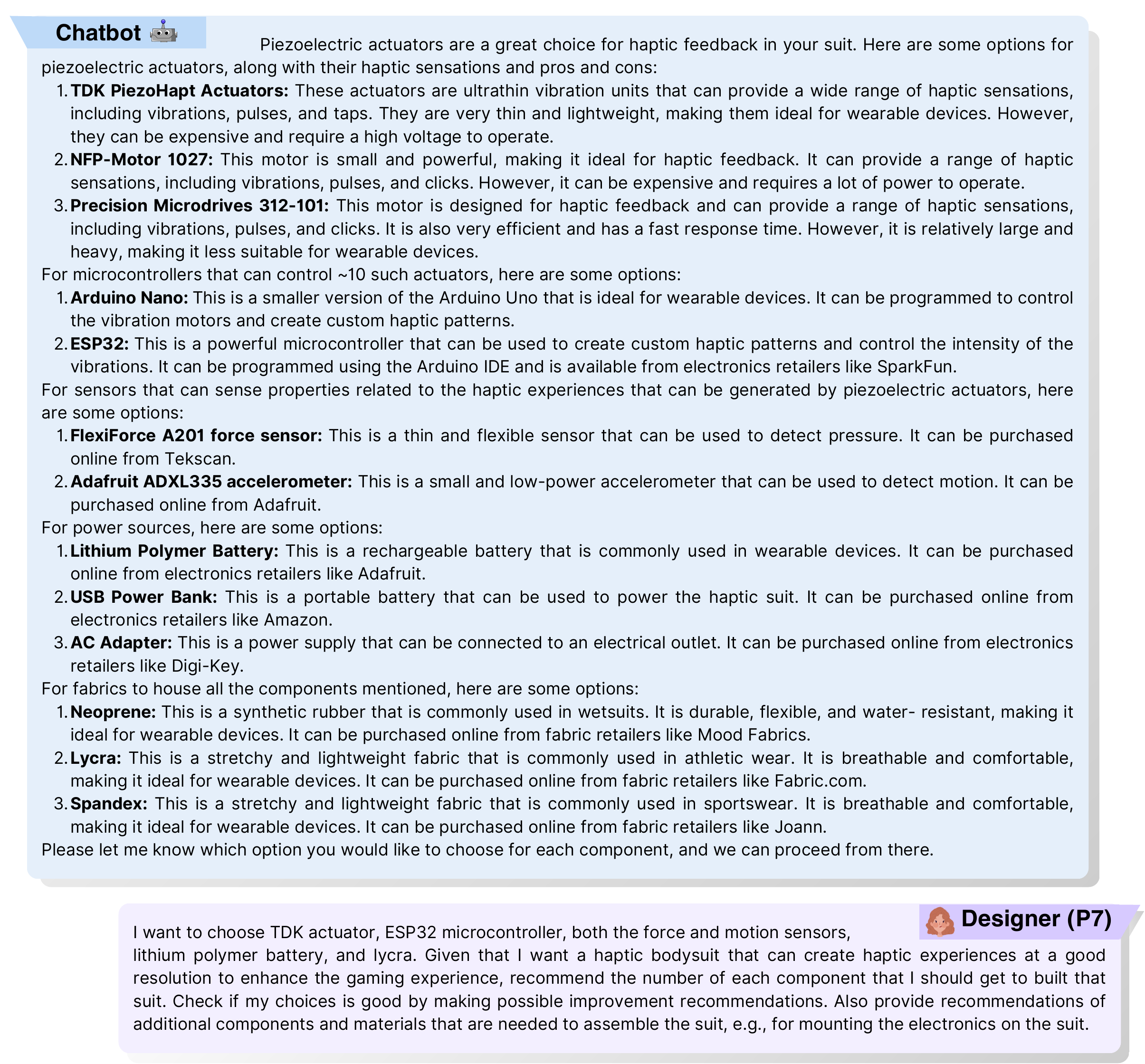}
    \caption{Third part of the input and output developed by P7 to select components for a haptic suit.}
    \label{fig:component_selection3}
\end{figure*}

\begin{figure*}[h!]
    \centering
    \includegraphics[width=0.9\textwidth]{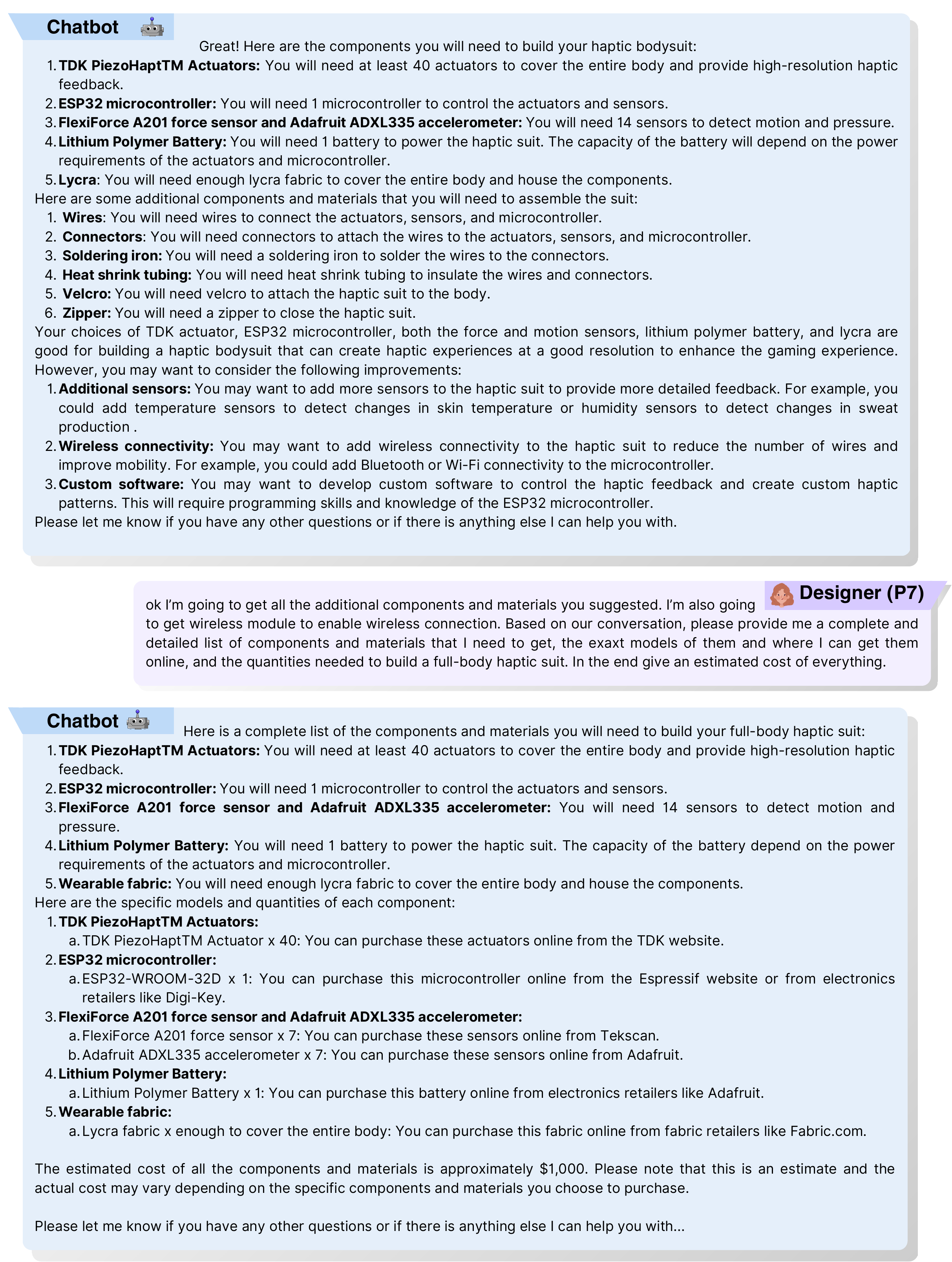}
    \caption{Final part of the input and output developed by P7 to select components for a haptic suit. Finally, P7 wraps up the selection procedure.}
    \label{fig:component_selection4}
\end{figure*}

\clearpage
\section{Designing a Controlled Experiment}\label{app:experiment_design}

\begin{figure*}[h!]
    \centering
    \includegraphics[width=0.95\textwidth]{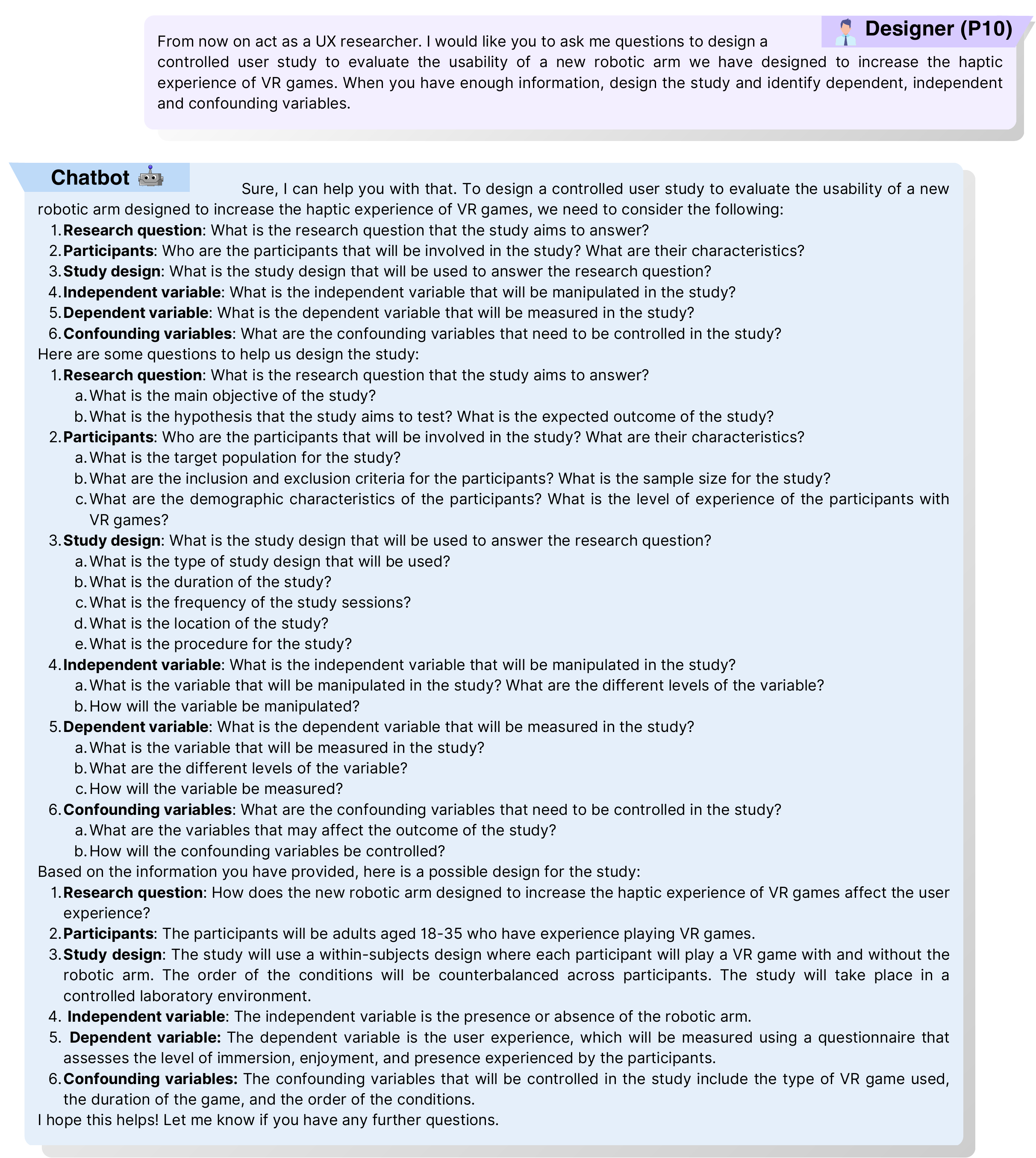}
    \caption{First part of the detailed input and output developed by P10 to design a controlled experiment. P10 deployed a flipped interaction pattern. Copilot replied with many questions.}
    \label{fig:experiment_design1}
\end{figure*}

\begin{figure*}[h!]
    \centering
    \includegraphics[width=0.95\textwidth]{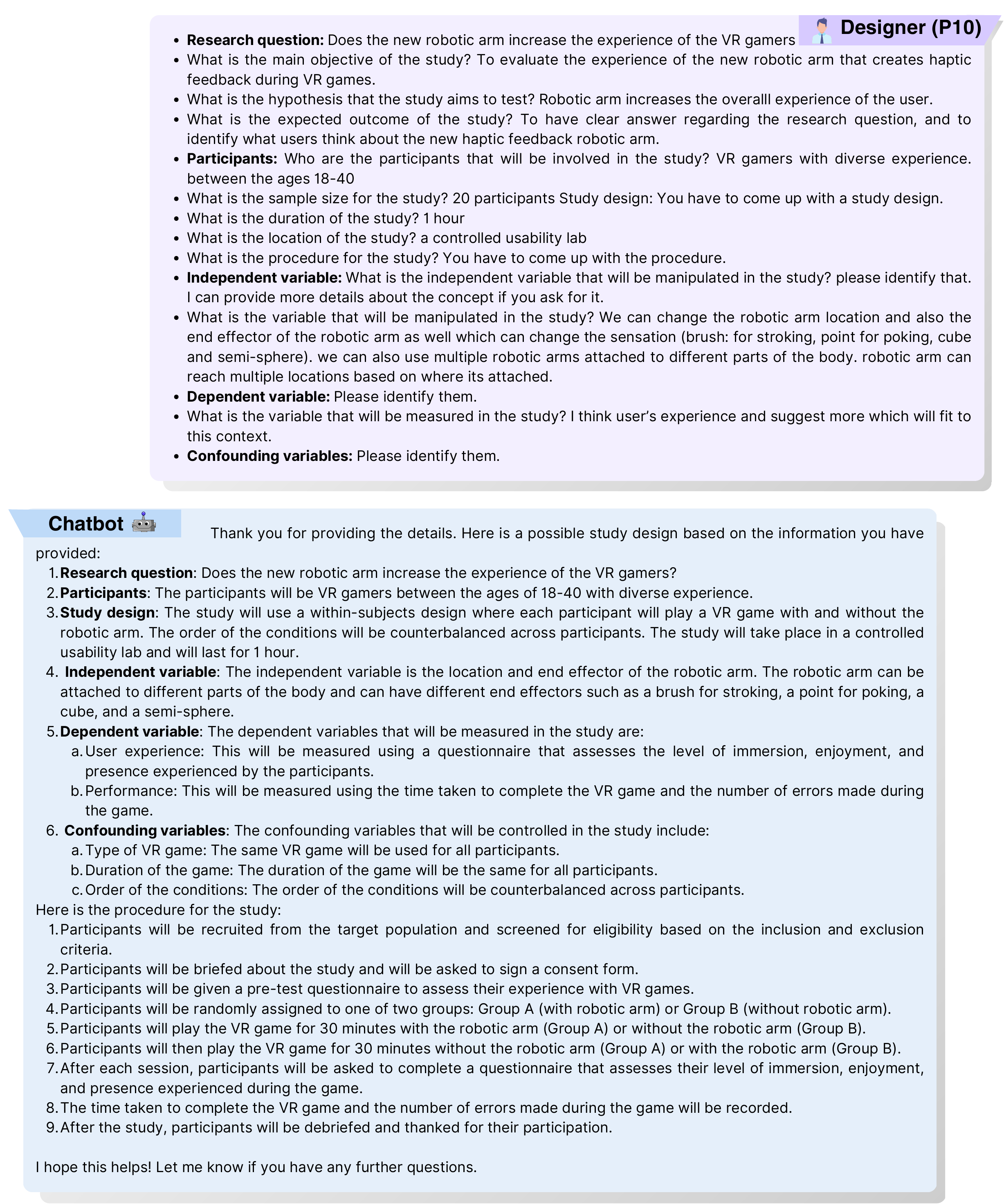}
    \caption{Second part of the input and output developed by P10 to design a controlled experiment. P10 selectively answered the questions and Copilot suggested a refined study design.}
    \label{fig:experiment_design2}
\end{figure*}

\clearpage
\section{Reviewing a Controlled Experiment Design}\label{app:evaluate_experiment}

\begin{figure*}[h!]
    \centering
    \includegraphics[width=0.85\textwidth]{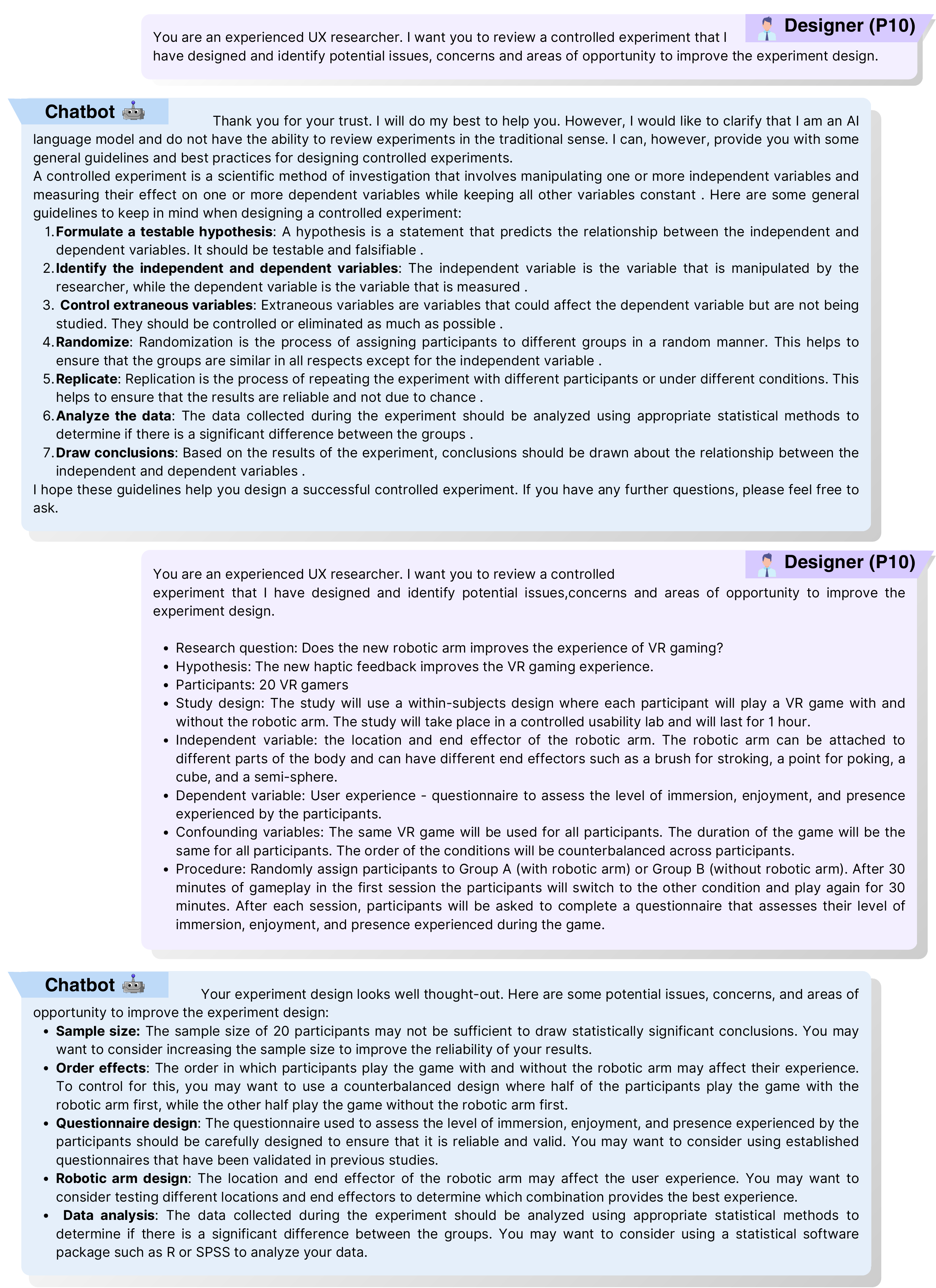}
    \caption{Detailed input and output developed by P10 to review a design of a controlled study.}
    \label{fig:evaluate_experiment}
\end{figure*}

\clearpage
\section{Improving a Controlled Experiment Design After a Pilot Study}\label{app:modify_experiment}

\begin{figure*}[h!]
    \centering
    \includegraphics[width=0.8\textwidth]{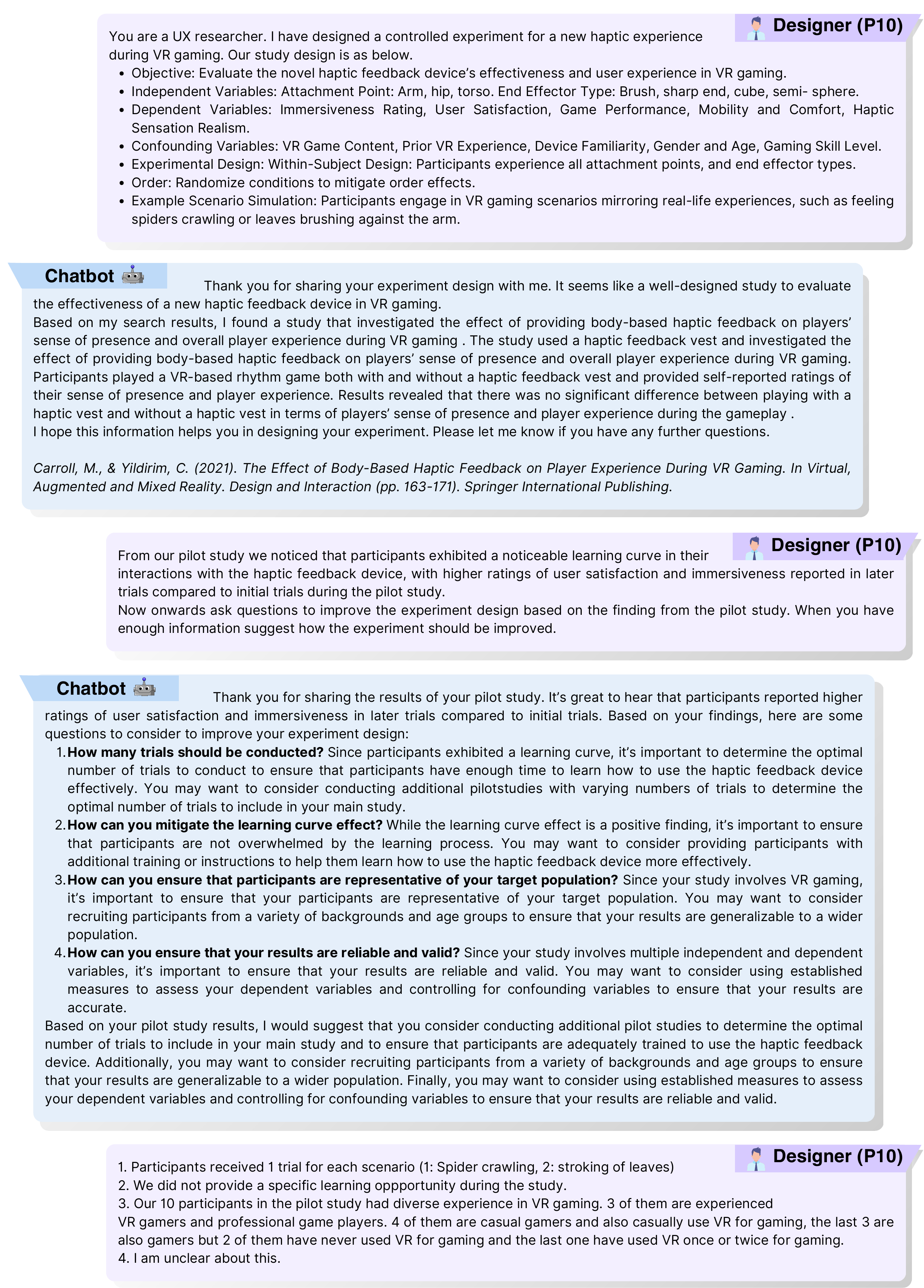}
    \caption{First part of the detailed input and output developed by P10 to modify the design of a controlled experiment based on insights from a pilot study. P10 deployed a flipped interaction pattern.}
    \label{fig:experiment_experiment1}
\end{figure*}

\begin{figure*}[h!]
    \centering
    \includegraphics[width=0.9\textwidth]{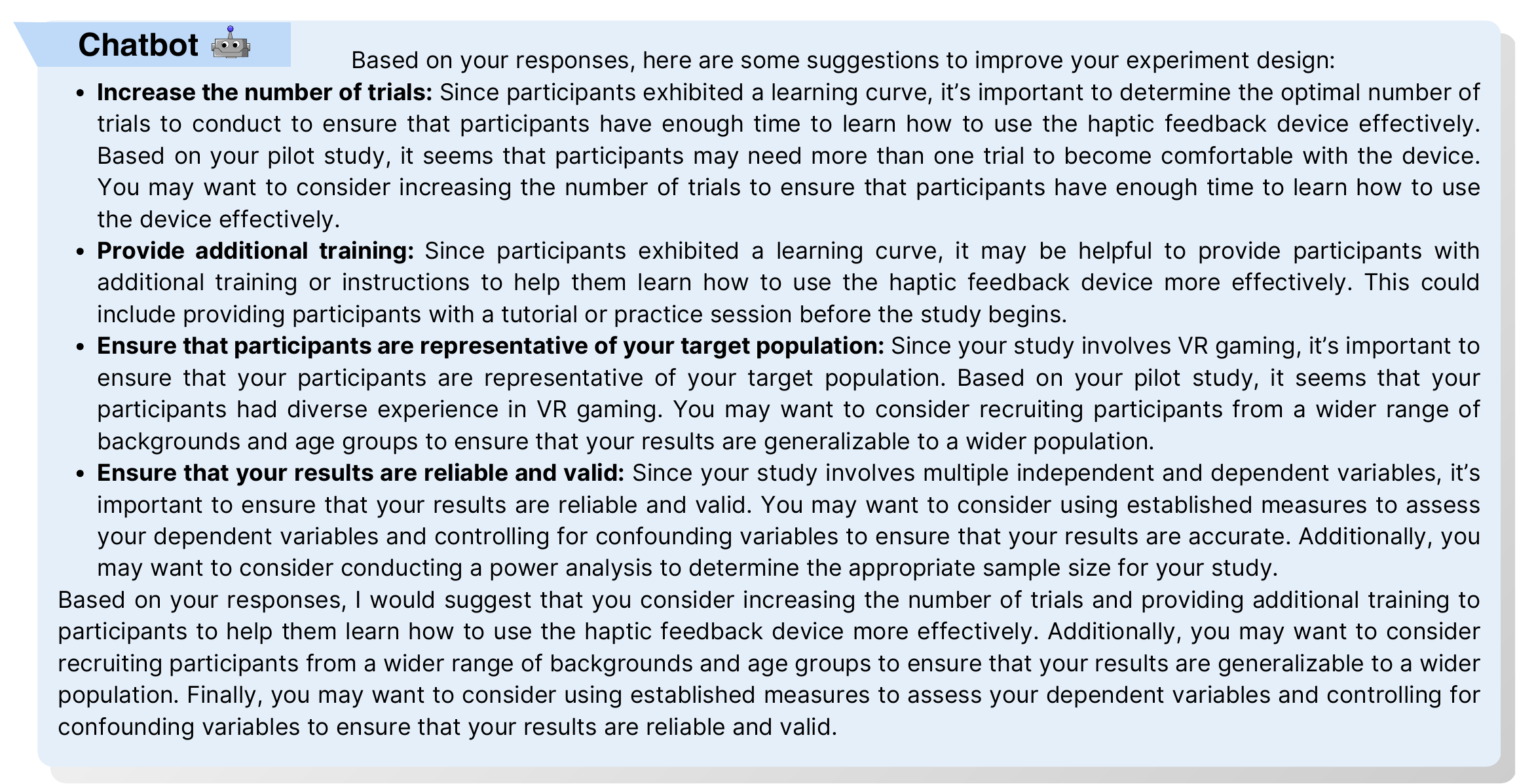}
    \caption{Second part of the input and output developed by P10. Copilot generated suggestions how the design of the controlled experiment could be improved.}
    \label{fig:modify_experiment2}
\end{figure*}
\end{document}